\documentclass[AER]{AEA}
\usepackage{natbib}
\usepackage{graphicx}
\usepackage{lscape}
\usepackage{booktabs}
\usepackage[sidewaysfigure]{rotating}
\usepackage{hyperref}
\usepackage{setspace}
\usepackage{subfigure}
\usepackage{float}
\usepackage{amsmath}

\usepackage[a4paper,bindingoffset=0.2in,%
left=1in,right=1in,top=1in,bottom=1in,%
footskip=.25in]{geometry}
\hypersetup{
	pdftitle={Monetary Policy and Wealth Inequality in Great Britain: Assessing the role of unconventional policies for a decade of household data},    % title
	pdfauthor={Anastasios Evgenidis and Apostolos},     % author
	colorlinks=true,                % makes links a coloured word, rather than putting coloured boxes around them
	linkcolor=blue,                % makes internal links black
	citecolor=blue,                % makes citations black
	urlcolor=blue,                 % color of external links
	breaklinks=true                 % allow links to be brocken across multiple lines
}

\draftSpacing{1.5}
\begin{document}

\openup .5em

\title{Monetary Policy and Wealth Inequalities in Great Britain: Assessing the role of unconventional policies for a decade of household data}
\shortTitle{Monetary Policy and Wealth Inequalities in the Great Britain}
\author{Anastasios Evgenidis and Apostolos Fasianos\thanks{
Evgenidis: Newcastle University Business School, Fasianos: Council of Economic Advisors, Hellenic Ministry of Finance, a.fasianos@minfin.gr. The authors would like to thank Leo Krippner and Andreas Zervas, for their useful advice, as well as participants in the Applied Macroeconomic and Empirical Finance Conference in Thessaloniki 2019 for their helpful comments. We are thankful to the UK Data Service for updating the WAS data with relevant information that allowed us to construct the inequality series employed in this study. Without their help our analysis would not be feasible. All remaining errors are ours. The views expressed in the paper are those of the authors and do not necessarily reflect those of the Council of Economic Advisors, Hellenic Ministry of Finance. }}
\date{\today}
\pubMonth{December}
\pubYear{2019}
\pubVolume{}
\pubIssue{}

\begin{abstract}
	This paper explores whether unconventional monetary policy operations have redistributive effects on household wealth. Drawing on household balance sheet data from the Wealth and Asset Survey, we construct monthly time series indicators on the distribution of different asset types held by British households for the period that the monetary policy switched as the policy rate reached the zero lower bound (2006-2016). Using this series, we estimate the response of wealth inequalities on monetary policy, taking into account the effect of unconventional policies conducted by the Bank of England in response to the Global Financial Crisis. Our evidence reveals that unconventional monetary policy shocks have significant long-lasting effects on wealth inequality: an expansionary monetary policy in the form of asset purchases raises wealth inequality across households, as measured by their Gini coefficients of net wealth, housing wealth, and financial wealth. The evidence of our analysis helps to raise awareness of central bankers about the redistributive effects of their monetary policy decisions. 
\\

JEL\ Classification: D31, E21, E52,  H31 \\

Word Count: 12,436

\end{abstract}

\maketitle

%1. Hook: Attract the reader’s interest by telling them that this paper relates to something interesting. What makes a topic interesting?

\section{Introduction}

\begin{figure}
\caption{Wealth Inequality in Great Britain (2006-2016)}
	\hfill
	\subfigure[Shares of Net Wealth for different household deciles]{\includegraphics[width=.45\textwidth, height = 5.5cm]{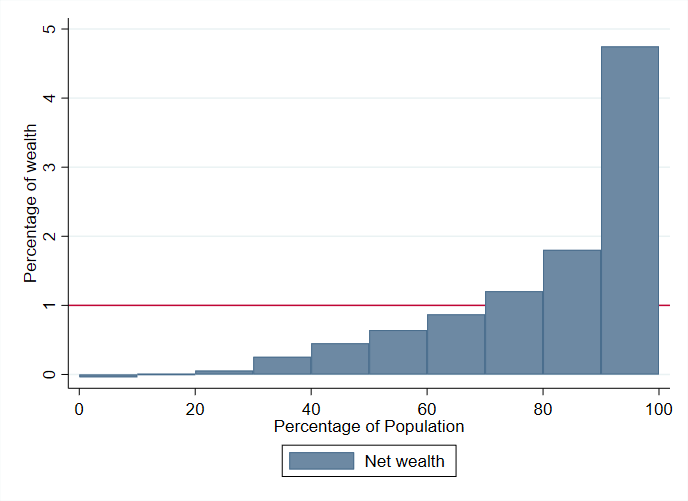}}
	\label{fig:wealth_shares}
%	\hfill
%	\subfigure[Cummulative Distribution of total net wealth]{\includegraphics[width=.5\textwidth, height = 5.7cm]{nw_cum_dist_yr.pdf}}
%	\hfill
%	\subfigure[Cummulative Distribution of wealth]{\includegraphics[width=.45\textwidth, height = 5.7cm]{lorenz_yr.pdf}}
%	\hfill
	\subfigure[Evolution of Net Wealth Gini]{\includegraphics[width=.45\textwidth, height = 5.7cm]{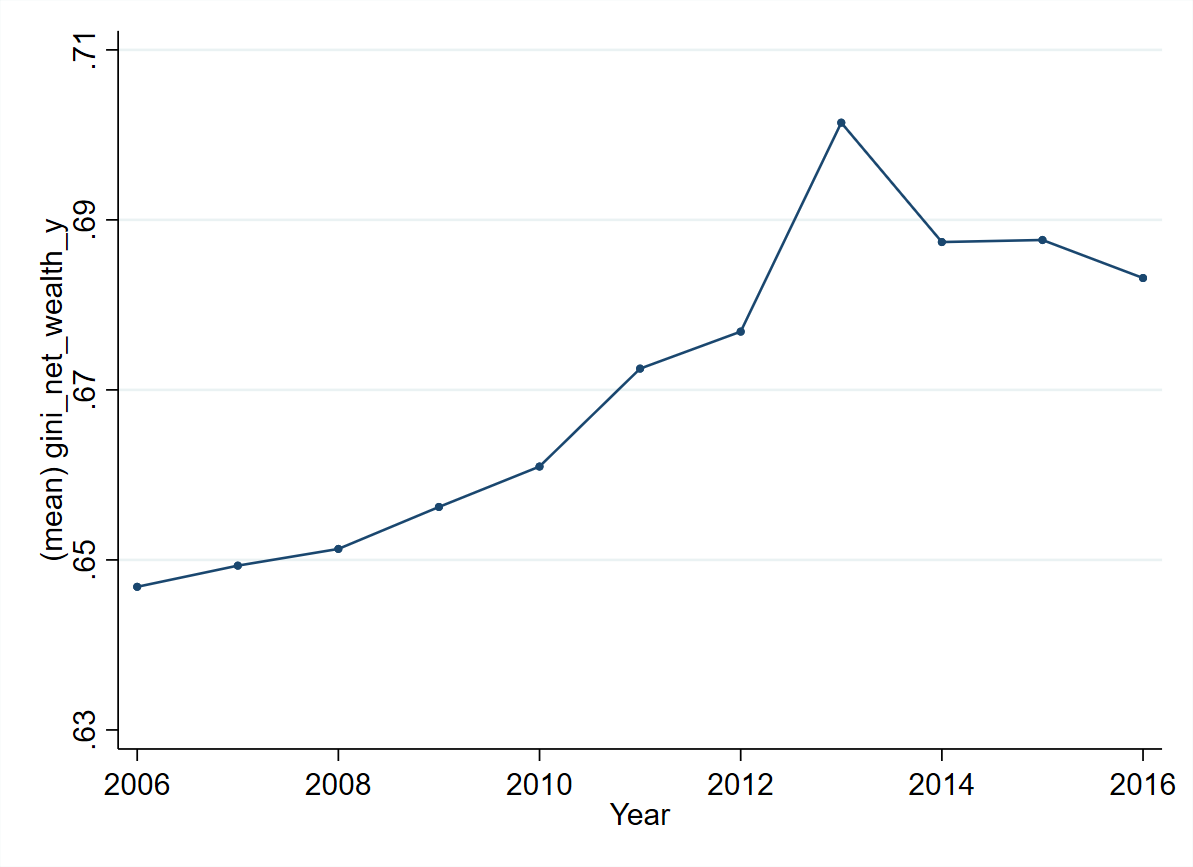}}
	\label{fig:wealth_evol}

\begin{figurenotes}[Source]
	Authors' estimations based on WAS Data \citep{ons2019was}.
\end{figurenotes}
\end{figure}

The 2007-08 financial crisis led to a profound shift in monetary policy. As policy rates reached the zero-lower bound (ZLB), central banks employed unconventional policies aimed at boosting nominal spending, increasing the liquidity of the financial system, and reaching their inflation targets. Unconventional policies played a significant role in alleviating the impact of the Global Financial Crisis (GFC), but also triggered  policy concerns that QE measures can have large effects on economic inequalities \citep{CASIRAGHI2018215, colciago2019}. Over the last 15 years, Great Britain  witnessed increasing levels of wealth inequality \citep{alvaredo2018top}. The share of the net wealth of the richest 10\% of the population accounts for almost 50\% of the country's total net wealth (Figure \ref{fig:wealth_shares}a), while overall wealth inequality increased by more than 4\% from 2006 to 2016 (Figure 	\ref{fig:wealth_evol}b). Although there is by now a growing literature exploring the relationship between monetary policy and income inequality,\footnote{See for example, \cite{mumtaz2017impact, guerello2018conventional} and \cite{colciago2019} for a literature review on the subject.} little research has focused on the impact of monetary policy on wealth inequalities.\footnote{Notable exceptions include \cite{saiki2014does, coibion2017innocent, auclert2017monetary}} Yet, monetary policy, and particularly unconventional measures, can influence household wealth shares by re-valuating and re-balancing their portfolios through different transmission channels \citep{ofarrell2017,colciago2019}. Understanding the impact of monetary policy on wealth inequalities is important for policy-makers because wealth is associated with households' financial health, it reflects households' future well-being, it is associated with political power  \citep{cowell2015wealth}, but also because wealth disparities imply heterogeneous consumption elasticities which can function as a transmission mechanism of monetary policy themselves \citep{kaplan2018, auclert2017monetary, arrondel2019does}.

This paper studies whether and how the unconventional monetary policy (UMP) operations implemented by the BoE affected financial and housing wealth inequalities in Great Britain for the period 2006-2016. It contributes to the empirical literature investigating the distributional effects of monetary policy on wealth inequality as follows: 

First, this is the first study investigating the distributional effects of monetary policy on wealth inequality for the UK by using low-level household balance sheet data at a relatively high frequency. Investigating wealth inequalities in relation to monetary policy has received much less attention than income inequalities in the literature (with some notable exceptions including \cite{Adam2016,coibion2017innocent,CASIRAGHI2018215,lenza2018does,hohberger2019distributional}. This gap arises not due to lack of research interest or policy relevance, but mainly due to serious data limitations. Reliable data reflecting the household portfolio distribution across the entire population distribution are themselves scarce; to say nothing about the availability of data reflecting their short-term dynamics, which are extremely valuable to investigate portfolio responses to monetary policy developments. Against this background, we draw on the Wealth and Asset Survey, a large sample survey on household finances in which individual responses are balanced proportionately over time and geography to construct monthly indices of net wealth inequalities for the period 2006 - 2016. In this way, our paper contributes to the broader wealth inequalities literature by providing a range of unique time-series of financial and housing wealth inequality estimates for the period preceding and following the Global Financial Crisis. Our ten-year estimated monthly series can be used by researchers aiming to assess the impact of other policies on the evolution of wealth inequalities.

Second, most of the existing empirical literature uses micro-simulation exercises \citep{doepke2006inflation, Adam2016,pugh2018distributional,lenza2018does} that typically provide information on household wealth or income distribution under policy scenarios which deterministically affect one or more of the distribution's components. These techniques fail to capture the underlying economic mechanisms being at play, especially when the policy effects are indirect or take time to realize. Even when micro-simulations allow for a stochastic component, they usually have only a few data points that prevent a long-run dynamic exploration of the relationship \citep{colciago2019}. To the best of our knowledge, this is the first paper that applies structural vector autoregressive (SVAR) models to investigate the impact of UMP shocks on wealth inequality. The available evidence on the use of multivariate time series models to examine the effect of monetary policy on inequality is limited with the exception of the studies of \cite{saiki2014does}, \cite{ guerello2018conventional}, and  \cite{inui2017effects} who yet focus on the impact of monetary policy on income inequality or apply local projections. Instead, we adopt Bayesian methods to estimate our models. Recently, Bayesian VAR methodology has become a relevant tool for evaluation of the effects of conventional monetary policy shocks (see, for instance, \cite{bandura2010, gali2015effects,  mumtaz2019dynamic}). The Bayesian approach offers a solution to the curse of dimensionality problem by shrinking the parameters via the imposition of priors and, as discussed in \cite{koop2010bayesian}, this approach offers a convenient method to estimate precise error bands for impulse responses.

Third, our paper builds on the literature of monetary policy channels redistribution and the different impacts of conventional and unconventional monetary policy. In contrast to most empirical literature on wealth inequality which focuses on the financial asset inflation channel (see, for example, \cite{inui2017effects,ofarrell2017, hohberger2019distributional}), we specify the broader portfolio rebalancing mechanism being at play under periods of UMP. In line with the relevant literature \citep{joyce2012}, our model controls both for the effect of increases in financial asset prices, but also for decreases in corporate bond yields, and the secondary effect of increases in housing asset prices. Furthermore, we investigate the savings redistribution channel from savers to borrowers, that functions on the opposite direction to the portfolio composition channel through drops in market net borrowing rates. Thus, our approach exploring the effects of the different channels simultaneously, also adds to the literature as most empirical studies analyze each distributional channel in isolation \citep{colciago2019}.

%The key transmission channel from UMP to wealth inequality is via the portfolio composition channel. When conventional monetary policy reaches the ZLB and UMP is employed, asset prices rise disproportionately compared to other economic variables, e.g., wages and employment. Inflated asset prices benefit primarily wealthier households that hold a larger portion of these assets and those households, in turn, receive greater capital gains and capital income.
%However, 

Fourth, to explore further the effects of UMP shocks on wealth inequality, we carry out counterfactual policy analysis that allows us to explicitly measure what would have happened to inequality had the BoE reversed its QE policy earlier. Therefore, our paper contributes to the related literature (see, \cite{bivens2015}) by providing three counterfactual policy experiments from two different VAR models, two conditional forecasts that are built on the main BVAR model used to analyze impulse responses, and a structural approach based on a time varying VAR model. Furthermore, we compare the effects of conventional versus unconventional monetary policy on wealth inequality. To assess this, we use a Bayesian threshold VAR that allows us to endogenously identify two different monetary policy regimes, i.e. zero lower bound (ZLB) and non-ZLB states and estimate the asymmetric impact of monetary policy shocks on wealth inequalities across the two regimes.  

Two sets of results emerge from our analysis that shed light on theories linking monetary policy and wealth inequality. 
First, impulse responses analysis suggests that unconventional monetary policy shocks elicit significant long-lasting effects on wealth inequality: an expansionary monetary policy in the form of asset purchases raises wealth inequality across households as measured by their Gini coefficients of total, housing, and financial net wealth as well as across wealth quantiles. In numbers, the shock is estimated to increase the Gini coefficient of total wealth by almost 0.03\% or 0.06 original units one year after the change in policy. In addition, forecast error variance decomposition shows that the contribution of UMP shocks to fluctuations in the wealth Gini is around 11\% at the first year horizon, suggesting that UMP measures did play an important role in the widening of the wealth gap over the forecasting period. The main results remain invariant to alternative specifications of the BVAR. We test their sensitivity to alternative specifications of the benchmark BVAR including different measures of monetary policy in the ZLB and measures of wealth inequality, different identification methods, and a medium-scale BVAR model that expands the information set in the benchmark case. Our results remain robust to all specifications.

Second, we find that the portfolio rebalancing channel, via the effect of elevated financial asset prices and lower corporate bond yields, plays an important role on the widening wealth gap. A further counter-intuitive finding is that the redistributive effect of UMP from poorer to richer households is less prominent but still present through the house prices channel. This result echoes the home-ownership structure prevailing in Great Britain, and contradicts most empirical studies, which predict house price increases to offset the regressive outcomes of financial asset inflation. On the contrary, UMP shocks lead to a fall in wealth inequality via the savings redistribution channel, indicating that falls in the net borrowing rate do redistribute wealth from savers to borrowers. Yet, the savings redistribution channel is not strong enough to offset the upward pressures on inequality elicited by the other two channels. Last, counterfactual policy experiments corroborate the results of the impulse response analysis suggesting that QE did aggravate wealth inequality.

From a policy-making perspective, our results suggest that although UMP measures have proven to be a powerful monetary instrument to boost liquidity and investment when the ZLB is binding, they need to be qualified by acknowledging their undesirable side effects, namely widening wealth disparities. Thus, our evidence informs policymakers in designing monetary policy by taking into account its impact on wealth inequalities.

The remainder of the paper is organized as follows. Section \ref{sec:theory} discusses the empirical background and the underlying theoretical intuitions on how monetary policy affects wealth inequality. The section identifies the key transmission channels behind the relationship, linking them to our empirical strategy.  Section \ref{sec:data} presents the various data-sets used in this study including the Wealth and Asset Survey, the short-term shadow rates (SSRs), used to identify monetary policy changes under the ZLB. Additionally, it provides information on wealth inequality summary statistics for the period in question (2006-2016) and presents the construction of the inequality series. Section \ref{sec:model} discusses empirical strategy and identifying assumptions behind the main Bayesian Var model employed and its variants. Section \ref{sec:results} presents the main results and explores alternate mechanisms that may be driving the relationship between monetary policy and wealth inequality. In addition, it provides a number of robustness checks for the main model results using different identification strategies for each of the two key variables. Section \ref{sec:conclusion} concludes.

\section{Distributional impacts of monetary policy and household heterogeneity: theoretical intuitions of the channels}
\label{sec:theory}

\subsection{Empirical Background}

Most empirical studies in the relevant literature focus on the interaction between conventional monetary policy and income inequality. Recently, an increasing amount of studies have begun investigating the influence of unconventional monetary policies on economic disparities. The first study that explored this relationship was performed by \cite{saiki2014does} comparing conventional monetary policy measures with the QE implementation with respect to their impact on income inequality in Japan. By applying a vector auto regression (VAR) on survey data, they show that monetary policy worsened income inequality, once the Bank of Japan (BoJ) initiated open market operations. For the case of the US, \cite{bivens2015} compares UMP to a fiscal policy counterfactual scenario, and suggests that expansionary monetary policy reduces inequality by stimulating income, increasing employment, and wages, which primarily benefit the poorer segments of the distribution. Similarly, \cite{mumtaz2017impact}, analysing quarterly UK data for the period from 1969 to 2012, find that contractionary monetary policy explains around 10-20\% of the variation in income inequality, as measured by the Gini coefficient of income, consumption, and wages. \cite{guerello2018conventional} disentangles the effects of conventional monetary policy (CMP) and UMP measures on the income distribution. Her measure of income inequality is constructed from the monthly Consumer Survey of the European Commission, which provides qualitative answers on a five-option ordinal scale. Her evidence from a VAR model suggest that CMP measures have negligible regressive effect on the income distribution. 

Much fewer studies looked at the relationship between monetary policy and wealth inequalities, particularly due to data limitations. Most of these studies focus on the portfolio composition channel and mainly do so using micro-simulation exercises and local projections to produce impulse responses. For instance, \cite{Adam2016} using the HFCS dataset, find that during the post-crisis period, Eurozone countries experienced an overall decrease in net wealth inequality, with increases in house prices counteracting the increase in wealth inequality caused by increases in equity prices. Their results vary depending on the relative distribution of housing and financial assets across countries. By contrast, \cite{domanski2016wealth} examine six advanced economies and find that net wealth inequality has risen since the financial crisis, with increases in equity prices outweighing house price increases. \cite{CASIRAGHI2018215} using quarterly data for Italian households simulate monetary policy impulses of household income and wealth statuses. Their evidence suggests a U-shaped response of net wealth on monetary policy, with poorer indebted households and asset-rich households mainly becoming better off, while those in the middle becoming worse off. \cite{pugh2018distributional} using a similar framework for a single wave of the WAS dataset for Great Britain, suggest that the overall effect of monetary policy on standard relative measures of income and wealth inequality has been small. In a similar fashion, \cite{lenza2018does} using the HFCS dataset, simulate short-run effects of UMP by the ECB on wealth and income inequalities for four euro-area countries, accounting for both employment and portfolio composition channels. An exception from microsimulation approaches is \cite{albert2018effects} who estimate a SVAR for income and wealth inequality for US and Eurozone. However, the use of the monetary base in this study as a proxy of QE is a poor identification strategy, as UMP implies additional pass-through channels, often not reflected in the monetary base (see \cite{joyce2012} for a relevant discussion). As pointed by \cite{colciago2019}, most of these approaches are not well equipped for capturing the multiple distributional dimensions of monetary policy as reflected in theoretical models.

\subsection{Wealth Inequality and Monetary Policy: Theoretical underpinnings}

The transmission of UMP relies more on wealth effects than that of CMP. Table \ref{tab: channels} summarizes the effects and the mechanisms of the main channels identified in the literature. The portfolio rebalancing channel is a crucial transmitter of UMP to the financial system and the economy. Under imperfect substitutability\footnote{ Imperfect substitutability may arise due to any asset feature, including convention and regulation. For instance, banks may be required to hold a certain amount of government bonds because they are risk free.} of assets, the announcement\footnote{Interestingly, UMP shocks can influence domestic equity returns and volatility, depending on the Bank’s information dissemination strategies \citep{chortareas2019}}.  of asset purchases by the central bank, can affect the composition of portfolios by influencing their relative supplies. If the assets purchased are not perfect substitutes for money, then the sellers of these assets rebalance their portfolios by buying other assets which are better substitutes. Hence, the price of the substitutes will increase to the point where the market for assets and money reaches an equilibrium. Increased asset prices imply lower yields, lower borrowing costs, and higher spending through wealth effects on consumption \citep{tobin1961, brunner1973mr, joyce2012}. As far as wealth inequality is concerned, the increases in asset prices will result in capital gains, that benefit particularly wealthier households who are the ones holding the bulk these assets across the wealth distribution \citep{coibion2017innocent,inui2017effects,hohberger2019distributional}.\footnote{An interesting extension of portfolio composition channels to periods of CMP has been proposed by \cite{BAGCHI201923} who theorize about the presence of Cantillon effects in the relationship \citep{cantillon1755essay}. When central banks increase money supply, the pool of banks' reserves widens, and new money is available for lending. Although the first to be granted those loans are not necessarily wealthier, they are able to bid up the price of capital goods. In effect, wealth inequality raises as banks channel new funds into the investments of the wealthier who have better access to finance and thus increase their purchasing power. By contrast, the poor with lower access to finance receive new funds only after its purchasing power has been eroded by inflation.}

Other transmission channels rely on savings redistribution mechanisms and include the effect of the borrowing/deposit rates and the unexpected inflation channel. Under expansionary monetary policy, either CMP or UMP, indebted households standing on the bottom of the wealth distribution, gain by experiencing a reduction in their interest payments on debt, while savers, standing on the top of wealth distribution, lose by experiencing lower returns \citep{pugh2018distributional, colciago2019}. In a similar manner, unexpected inflation favors debtors in the lower-middle of the wealth distribution who experience falls in the nominal value of their liabilities, but adversely affects richer households who invest in long-term bonds and are net lenders \citep{doepke2006inflation, meh2010aggregate}. 

It is worth mentioning that household heterogeneity is key in the transmission of redistributional wealth effects through both the portfolio rebalancing and the savings redistribution channels. This is because both asset inflationary and net interest effects caused by monetary policy shocks will ultimately depend on the household portfolio composition and the loan/deposits contract structure of a country's population.

\begin{table}[htbp]
  \centering
  \caption{Monetary policy and wealth inequality : Transmission Channels}
   \label{tab: channels}
             \begin{tabular}{l@{\hskip .4in}c@{\hskip 0.3in}c}

    Channel \& Monetary Policy & Effect &  Mechanism  \\
    \toprule
    \\
Portfolio rebalancing/ & Financial asset prices rise &  Compositional /  \\
Portfolio composition  & Short term bond yields fall &  Heterogeneity  \\
 (UMP) & Housing asset prices rise &   \\
 \\
 \hline
 \\
Savings  & Interest payments  & Compositional/   \\
 redistribution  &  and deposit rates  & Heterogeneity  \\
 (CMP/UMP) &  on trackers fall &   \\
 \\
 \hline
 \\
 & Falls (rises) in the  &  Inflationary \\
Unexpected inflation  (CMP*) &  nominal value  &  Expectations \\
  &   of debts (deposits) &    \\
\\
    \bottomrule
    \end{tabular}
  \label{components}%
  \begin{figurenotes}[*]
	Although there is no theoretical reason to assume that the unexpected inflation channel cannot function under periods of UMP, we did not highlight this possibility due to the lack of relevant studies in the literature and no big inflationary impact of the QE measures as of yet. 
	\end{figurenotes}
\end{table}

Our empirical strategy is closely associated with the model of \cite{hohberger2019distributional} who study the distributional effects of both CMP and UMP by comparing the impulse responses of net financial wealth shares to expansionary monetary policy shocks, among other inequality indicators. The authors estimate an open economy two-agent DSGE model that distinguishes between two types of households: On the one hand, "financial investors" or non-liquidity constrained households, that hold financial assets (government bonds, firm equity, and foreign bonds) receiving capital income, wages, and transfers, and, on the other hand, "asset-poor" or liquidity constrained households receiving only wages and transfers. UMP is modeled as an expansion of the central bank balance sheet by purchasing long-term bonds from households. In line with the portfolio rebalancing mechanism described above, prices of imperfectly substitutable short-term financial assets increase, leading to a drop in household savings, a reduction in firm financing costs, and exchange rate depreciation. In turn, increases in value of the asset-rich net wealth imply increases in wealth inequality, as liquidity constrained households have zero financial wealth by construction.

The results from their estimated DSGE model suggest that monetary policy, under both CMP and UMP scenarios, increases net financial wealth inequality but its effects are short-lived. Immediately after the shock, lower interest rates lead to increases in the value of assets held by asset-rich households. In the medium term, however, UMP reduces the net financial wealth share of NLC households, a result driven by the shrinking of the household's holding of interest-bearing long-term bonds and the reduction in private sector savings for an extended period. 

Our empirical model deviates from the above described behaviour as follows: In our empirical estimations households hold housing in addition to financial assets. By overlaying the net financial and net housing wealth distribution, one could distinguish between at least three types of households. Asset-poor liquidity constrained households holding neither housing or financial wealth, home-owners with limited financial wealth, and asset-rich households with large stocks of both financial and housing wealth. 

\begin{figure}[htbp]
	\caption{Total net wealth and each component across the wealth distribution}
	\label{perc_fin_hous}
	\includegraphics[scale=0.55]{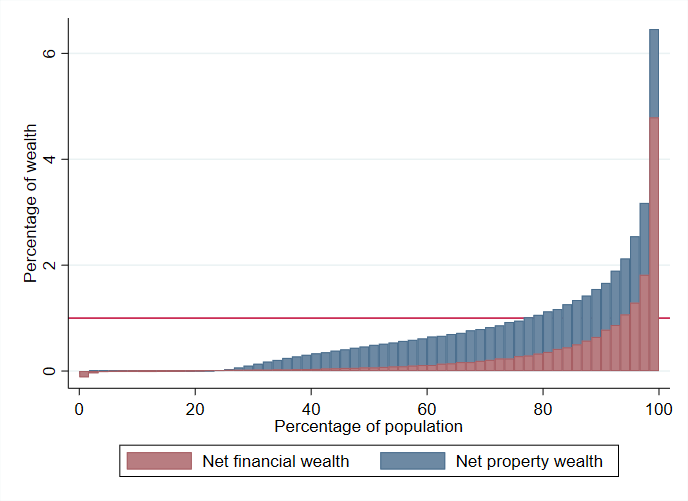} 
	\begin{figurenotes}[Note]
		Authors' estimations from Wealth and Asset Survey (2006 - 2016)
	\end{figurenotes}
\end{figure}

As shown in figure \ref{perc_fin_hous} net housing wealth is more equally distributed than net financial wealth with a greater concentration of direct financial assets among households at the top of the wealth distribution. Within financial wealth components, deposits and cash are more equally distributed while riskier investment assets such as stocks and bonds are disproportionately held by wealthier households. Two intuitions come from this chart: First, the disproportionate large holdings of financial by the top 20\% - 25\% of British households suggests that the portfolio composition channel of monetary policy may be at play and have an impact on inequality in a similar manner to the theoretical model described above. In our model we test this effect empirically by controlling for both rebalancing effects, namely asset price inflation and bond yields falls. 

Second, the much more proportionate, but still uneven, distribution of housing wealth may also imply effects of UMP on overall wealth inequality through the housing wealth channel, if present. Indeed, the effect of UMP on net housing inequality is less straightforward. This is because UMP affect the housing market in two ways: on the one hand, the purchase of mortgage-backed securities (MBS) by the central bank leads to large scale mortgage refinancing \citep{Krishnamurthy2011} which increases net wealth (mortgage debt) of liquidity constrained asset-poor households.\footnote{The purchase of MBS, however, is more relevant in the case of the US than the UK, where the FED bought this type of assets aiming to support a housing market where excessive sub-prime loan backed securities was one of the factors behind the crash.} On the other hand, UMP yields positive and persistent effects on house prices and residential investment \citep{rahal2016housing}, assets which are mainly held by the middle and the upper parts of the total wealth distribution. Henceforth, depending on the distribution of home-ownership, the value of housing assets, and the distribution of debts across households, wealth inequality may decelerate, stay intact or accelerate. The opposing evidence coming from various theoretical models assumptions suggests that the effects of monetary policy on wealth inequality is \textit{a priory} ambiguous. The effects of monetary policy-induced asset-price changes on net wealth inequality depend on the relative distributions of the assets and debts, which shape the different leverage and asset structures across the wealth distribution \citep{ofarrell2017}. As the rise in asset prices can either increase or reduce net wealth inequality and the evaluation of the effects deserve an empirical investigation. In the following sections, we describe our empirical strategy for assessing the impact on UMP for the case of Great Britain.

Last, our empirical strategy introduces the savings redistribution channel, by taking into account the effects of borrowing rate falls on the distribution between net savers and debtors. We expect the falls in interest rates to function at the opposite direction than the portfolio composition channel.

\section{Constructing monthly estimates of wealth inequality in the UK}
\label{sec:data}
\subsection{Data}

Most existing time-series analysis estimating the impact of monetary policy shocks on wealth inequality indices are based on aggregate macroeconomic indicators regressed against partly disaggregate inequality measures (e.g., Gini coefficients of income, wages, consumption, and rarely wealth). A caveat of those studies is the lack of reliable data, since surveys on household balance sheets are usually provided at a low frequency, typically every one or two years \citep{Alvaredo2016challenge, pugh2018distributional}. The focus of our analysis is on whether UMP shocks impact on the distribution of households' wealth components, real and financial assets and debts. Annual or bi-annual data on wealth would not allow us to assess short-term policy impacts. Instead, we require low level household balance sheet data held by households on a relatively high frequency, as well as information on the distribution of these items across wealth percentiles. To achieve our goal, we estimate changes in wealth inequality using information from the Wealth and Asset Survey (WAS) for the period 2006 - 2016 \citep{ons2019was}. WAS is a longitudinal household survey, in which  households in Great Britain\footnote{Excluding addresses north of the Caledonian Canal, the Scottish Islands and the Isles of Scilly.} are interviewed every two years. Approximately 30,000 households were interviewed in wave 1, 20,000 in wave 2, 21,000 in wave 3, 20,000 in Wave 4, and 18,000 in Wave 5.\footnote{For cost-efficiency reasons, only 57\% of the full sample of 30,595 responded to all components of their wealth in the first wave. Consequently, this sub-sample will be included in this analysis with regards to the first wave.}

An important feature of WAS is that interviews were allocated over the 24-month fieldwork period using systematic sampling with a random start point such that interviewees' addresses were balanced proportionately over time and across space. The proportional allocation across months and regions for all survey waves, allows us to break down the full sample population into monthly representative sample cohorts from the date the fieldwork commenced (July 2006) till its last released wave (June 2016). In total, we have around 100,000 observations of household net wealth for the entire period. This is a substantially large sample period for our purposes, as it covers the entire UMP implementation sample as well as a pre-crisis sample, when the BoE implemented traditional CMP. Each monthly cohort comprises a minimum of 800 households, a number large enough to provide unbiased inequality estimations using the Gini coefficient.\footnote{According to \cite{deltas2003small} the Gini can be biased downward in small samples, because for a given level of intrinsic inequality, a reduction in the sample size leads to reductions in apparent inequality as
measured by the Gini coefficient.} Furthermore, our constructed monthly cohorts also satisfy the theoretical requirements of a synthetic panel cohort, and, as such, can be seen as a monthly representative sample of the UK population \citep{Verbeek2008}. Taking advantage of the structure of WAS and the fact that interviews took place proportionately throughout the year, we follow \cite{cloyne2016household}, \cite{mumtaz2017impact}, and \cite{inui2017effects} and assign households to different months each year based on the date of survey interview. This feature allows us to calculate the measures of inequality at a monthly frequency. 

A constraint on providing reliable estimates of wealth inequality is the high non-response rates of the wealthiest segments of the distribution. For instance, \cite{vermeulen2018fat} compared the WAS estimates with those of the Forbes' Billionaires list assuming a Pareto distribution. His results suggest that WAS underestimates the top 1 per cent share of wealth by 1 to 5\%. The lack of reliable data representing the richest wealth shares partly explains the gap in the literature for UK estimates of wealth inequality despite increasing interest on the subject in the last ten years \citep{Alvaredo2016challenge, alvaredo2018top}. Ideally, the distribution of wealth would be estimated in relation to the external control totals for population, based on administrative personal wealth data \citep{Alvaredo2016challenge}. Unfortunately, in the UK there is no annual wealth tax so we could not apply such a control on our survey estimates.\footnote{Recent work by \cite{alvaredo2018top} addressed this caveat by imputing long-run estimates of wealth inequality since 1895 using the distribution of estates at death. However, their yearly estimates cannot be used for short-term analysis or for decomposing between real and financial wealth.} This is a limitation of our dataset and, therefore, we recommend a degree of caution in the interpretation of our results, due to possible under-representation of extremely rich households, and thus underestimation of the overall level of wealth inequality. Yet, another feature of the WAS is that it over-samples the wealthy addresses and therefore has an improved representation of the right-skewed upper tail despite high non-response rates of the extremely rich.\footnote{In particular, the survey is constructed such that it over-samples wealthy addresses using a rate of 3 for the wealthiest addresses (ONS, Wealth and Asset Survey Report).} Furthermore, the central scope of this paper is to capture short-term effects of UMP on housing and financial wealth inequality changes, rather than measuring wealth inequality for the entire population \textit{per se}. 

The investigation of UMP effects on wealth inequality carries an additional identification problem \citep{colciago2019}. As the period of UMP overlap with the period of the ZLB, it is tricky to disentangle the impact of low interest rates, UMP, or the interaction of both. To identify UMP changes we use the short-term shadow rate (SSR) as a measure of the monetary policy stance extracted by modelling the term structure of the yield curve. To the best of our knowledge, only \cite{inui2017effects} follow a similar identification approach to UMP shocks on inequality for the case of Japan. The level and slope of the yield curve provides information on investors' perceptions and expectations with regards to future monetary policy actions and about the course of the interest rates. Certainly, the strongest advantage of SSRs is that they are a powerful instrument to predict the monetary policy stance at the zero lower bound (ZLB) with the data from the non-ZLB period in a fixed-parameter model \citep{bullard2012shadow, wu2016measuring,lombardi2018shadow}. An additional advantage of SSRs over other proxies of UMP measures, is that SSRs use information from the entire yield curve, including forward guidance, quantitative programmes, and their announcements.\footnote{Because financial markets are forward looking, the impact of asset purchases on bond yields is likely to occur when expectations of purchases are formed and not necessarily when the purchases are realized. For instance, \cite{joyce2012qe} show that the biggest impact of QE1 on bond yields took place when the purchases were initially announced.}  Consequently, SSRs can capture the overall effects of a given measure.\footnote{Another common proxy of UMP is estimations from central bank balance sheets, for instance, total Bank's assets. Balance sheet proxies reflect central bank's purchases at the time they were implemented, if implemented, rather than announcement impacts. An interesting example of the difference, was the ECB Open Market Operations (OMT) programme which implied large quantitative effects when announced, but it was never actually implemented. As a result, a significant UMP event would not be captured by a central bank balance sheet proxy. As BoE implemented all of their announcements, there should be no issue with using either proxy in our case. Henceforth, we run a robustness check using Bank's balance sheets in section \ref{robustness_checks} as well. } For instance, \cite{christensen2016modeling} compare the performance of the SSR model to a Gaussian Dynamic structure model on predicting bond yields at the ZLB, and show that the use of SSRs indicates higher in-sample fit, matches the compression in yield volatility, and delivers improved real-time short-rate forecasts than the standard model. In the present analysis we employ the two main SSR rates for the UK mostly cited in the literature, estimated by \cite{wu2016measuring} and by \cite{krippner2014measuring, claus2018asset}. The two series are presented in Figure \ref{fig:shadow_rate} in Appendix \ref{app:ssrs}.\footnote{The SSR series have been obtained from Cynthia Wu's website and can be accessed here: \url{https://sites.google.com/view/jingcynthiawu/shadow-rates} and Leo Krippner's website and can be accessed here: \url{https://www.rbnz.govt.nz/research-and-publications/research-programme/additional-research/measures-of-the-stance-of-united-states-monetary-policy/comparison-of-international-monetary-policy-measures}}

%include a short paragraph describing all the rest of the controls (macro vars etc)

\subsection{Wealth Inequality Estimates for the period 2006-2016}

The nature of wealth inequality often depends on the definition being adopted. Following the relevant literature, we consider current net wealth as the standard wealth concept in this analysis \citep{cowell2015wealth}. Put simply, current net wealth stands for the difference between assets and debts, depending on the type of wealth measured. It is defined as follows:

\begin{equation}
x = \mathop \sum \limits_{j = 1}^m {\pi _j}{A_j} - D{\text{}}
\end{equation}

Where $A>= 0$ is the amount held of asset type $j$, $\pi$ is its price and $D$ represents the call on those assets represented by debt.   

Financial assets (“Financial wealth” hereafter), include formal investments such as bank or building society current or saving accounts, investment vehicles such as Individual Savings Accounts (ISAs), endowments, stocks and shares, less any financial liabilities such as outstanding balances on credit cards, arrears on household bills, and loans from formal sources.\footnote{We exclude informal financial assets, e.g., money under the bed or loaned to family or friends), children’s assets and pension wealth, i) to reduce measurement error as formal financial assets and housing can be valuated by survey respondents with greater precision, and ii) because possible influence of monetary developments on informal assets and accumulated pension wealth is hard to capture in a structural model.} Non-financial assets (“housing wealth” hereafter) include self valuation of property owned, both main residence plus any other land or property owned in the UK or abroad; less the outstanding value of any loans or mortgages secured on these properties.\footnote{Self valuation tends to yield higher estimates of worth than most other property indicators may suggest – however, when assessing individuals’ opinions, it is this perceived worth that will drive the individuals concerned.}. We further exclude consumer durables such as automobiles and housing equipment. As suggested by \cite{wolff2009household}, although tangible assets carry a resale value, they can only accrue it by compromising current consumption. Other studies measuring wealth inequality from survey data use similar conventions (see, for example, \cite{cowell2017}) In line with previous studies on wealth distributions, the economic unit in this analysis is the household, where assets and debts are summed up for all household members. As there is currently no consensus in the literature on the need to equivalize wealth estimates, we do not adjust our wealth estimates for household composition.\footnote{As \cite{cowell2015wealth} point out, if one interprets wealth as potential future consumption, it is not current household structure that should matter, but future structure instead, perhaps after retirement when bequest motives are realized. As a result, it would only make sense to adjust wealth using a future family structure equalizer, which is of course not observable at present.} All households with missing values for assets or debts are dropped from the sample (less than 0.1\% of the total sample). Based on the above definition, we also produce the two main components of net wealth, namely housing and financial net wealth. Net housing wealth includes all housing assets held by the household minus mortgage debts. Net financial wealth includes all formal financial assets minus any type of formal debt secured on property. Lastly, we report a measure of augmented net wealth which includes all possible assets reported in the survey and official household wealth statistics, including informal assets and debts, all sorts of consumer durables and vehicles, collectables and valuables, as well as accumulated pension wealth.

Our main measure of wealth inequality is the Gini coefficient of household's net wealth ($G_{j,t}$), defined as follows:

\begin{equation}
G_{j,t} = \frac{{\sum\limits_{i = 1}^{n_{t}} {(2i - n_{t} - 1)x{'_{i,t}}} }}{{{n_{t}^2}\mu_{t} }}
\end{equation}

where $j$ is the type of assets that comprise each measure of net wealth (total net wealth, financial, housing), $t$ stands for the monthly sample that the interviewed household lies in, $i$ is the household’s rank order number, $n$ is the number of all households present in each month, $x_{i}$  is the household's $i$ net wealth value, and $\mu$ is the population average. The Gini coefficient ranges from 0 to 1, with 0 representing perfect equality and 1 representing perfect inequality. On top of being one of the most frequently used indicators of inequality, Gini measures of inequality are well defined for negative values and their properties are preserved \citep{cowell2015wealth}\footnote{Other Gini's desirable properties as an inequality measure include scale independence, population independence, symmetry, and the axiom of transfers}. Allowing for negativity of input values is crucial for exploring net wealth, as a portion of observations in our sample is lower than zero because debts may exceed assets at a given point in time. Gini coefficient estimates fed with negative wealth values implies that the index is not anymore upper bounded at 1, but can take any positive value. Regarding wealth estimates, there is no theoretical maximum for inequality since asset-free households can borrow infinitely to finance regressive transfers to rich ones \citep{cowell2017wealth}.

In the empirical model of this study, we consider three more common inequality indicators in our analysis and our robustness checks. Namely, we estimate the wealth percentiles (quantiles) showing how much share of the total net wealth is held by different shares of the population, the ratio of the wealth holdings of the top 20\% richer share of the population over the holdings of the 20\% poorer share of the population (20:20 dispersion ratio), and the coefficient of variation defined as the standard deviation divided by the mean of a given distribution.
 
 Table \ref{tab:wealth_desc} shows mean values for different assets and liabilities over the 10 year period examined in nominal Great British Pounds (GBP). Net wealth increased by 35\% over the entire period, while gross total wealth increased by 34\%. The mean value of all gross housing assets grew by 27\%, which is less than the overall percentage increase in gross wealth. The biggest gains stem from formal financial assets growth which accounts to 60\%. Augmented net wealth increased by more than 40\%, also more than main net wealth, implying a significant increase in pension wealth during this period. The mean value of total liabilities increased by 14\%, much less than the increase in total assets. Housing related liabilities increased by 11\% other financial liabilities increased by 42\%. 
 
We next turn to trends in wealth inequality using the aforementioned wealth concepts. Our estimates for the Gini coefficients of total net wealth, augmented net wealth, net financial wealth, and net housing wealth, are presented bi-annually in Table \ref{tab:ineq_desc}. On the basis of the WAS data, the Gini coefficient for net wealth increased substantially from 0.66 to 0.68, by 4.3\%. The same index peaked at the years between 2012 to 2015 at 0.69. The net financial wealth Gini stands around 0.9 for the entire period and is considerably larger than the net housing wealth Gini with is approximately 0.66. Yet, the latter indicated much larger increases of 4.34\% for the period in question, compared to slight increases of 0.5\% for the former. Similar conclusions are drawn by observing the estimates for quantile shares of the three wealth indices. The share of the richest 75\% held around 71\% in 2006/07 and almost 75\% of total net wealth in 2016 indicating an increase of almost 5\%. These increases are mostly attributed to housing wealth gains which increased by 5.7\% over the whole period, and less so to financial wealth gains which increased only slightly. Interestingly, the corresponding increases of the richest 10\% of the population are sharper, implying a large redistribution towards the upper segments of the net wealth distribution. 

\begin{figure}[htbp!]
	\caption{Net wealth Gini and its components from 2006 to 2016}
	\label{fig:gini_inequality}
	\includegraphics[scale = 0.28]{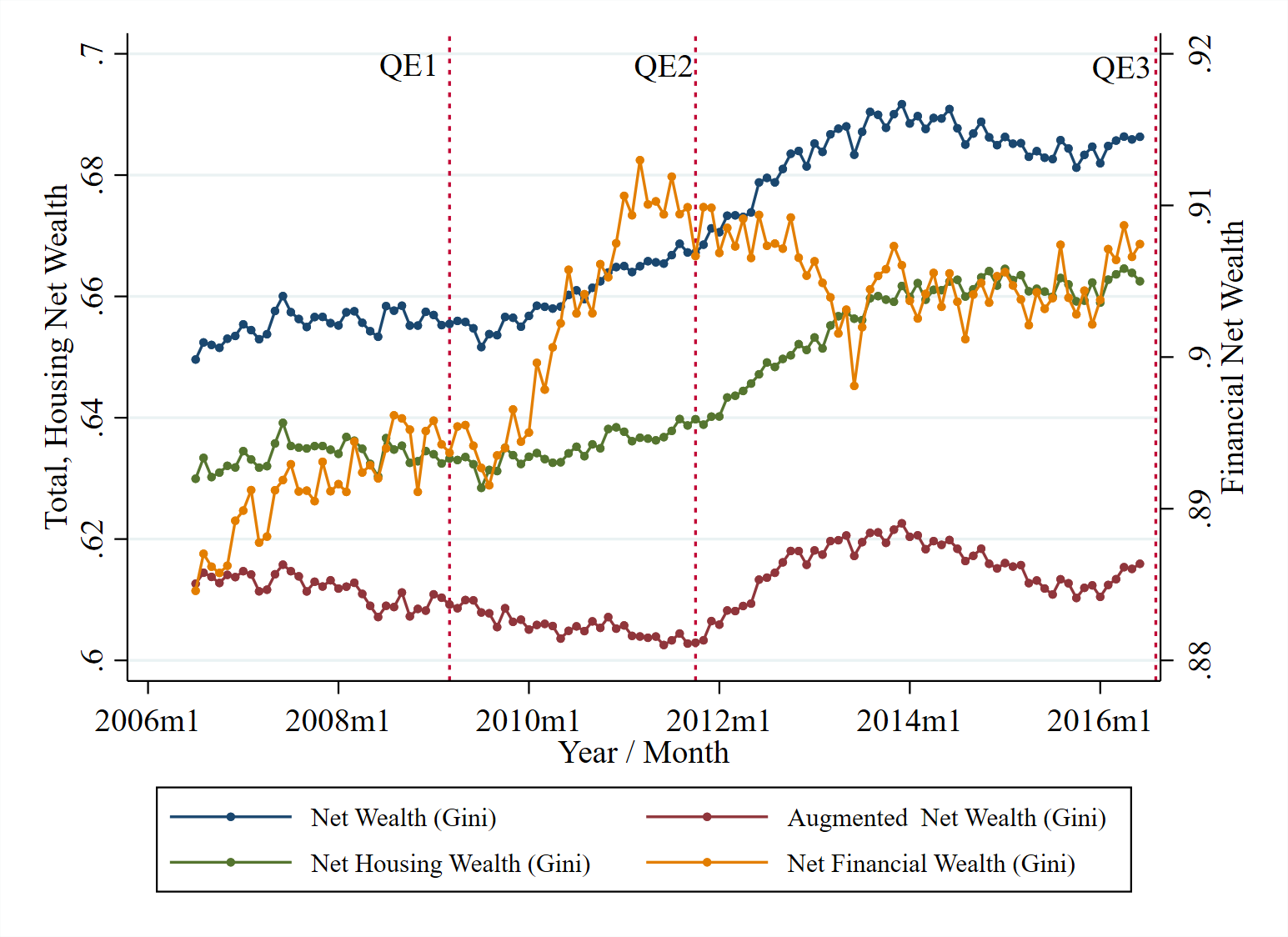} 
	\begin{figurenotes}[Source]
		Authors' estimations from Wealth and Asset Survey (2006 - 2016). 
	\end{figurenotes}
	\begin{figurenotes}[Note 1]
		Gini coefficients (six month-moving average). Net wealth refers to total household's assets minus liabilities, excluding pension wealth, tangible assets other than housing, and informal financial assets and debts. Augmented net wealth includes these assets. Net housing wealth includes all housing assets held by the household minus mortgage debts. Net financial wealth includes all formal financial assets minus any type of formal debt collateralized on property.
	\end{figurenotes}
	\begin{figurenotes}[Note 2]
		The red dotted lines depict the announcements of UMP by the BoE. 
	\end{figurenotes}
\end{figure}

Figure \ref{fig:gini_inequality} shows our estimations for the Gini coefficients of total net wealth, net financial wealth, net housing wealth, and augmented net wealth, as they evolve from 2006 to 2016 on a monthly basis, estimated according to the methodology described above. We also include the timeline of the various UMP rounds as announced by the BoE during the period in question. The first round of UMP (QE1) was announced on 5 March with an initial decision to purchase £75 billion of assets over 3 months financed by issuance of central bank reserves. The purchases increased to £200 billion over the next month and the scheme was completed in January 2010. The second round of UMP (QE2) through asset purchases begun in October 2011, in response to the impact of  Euro crisis. Additional £125 billion of	purchases completed in May 2012. Lastly, additional £50 billion announced in July 2012 and completed in November 2012. In August 2016, the BoE announced a third round of purchases (QE3) amounting to GBP 60bn of UK government bonds and £10bn of corporate bonds, to address uncertainty over Brexit and concerns about productivity and economic growth.	Being equipped with these series, in the next section we set-up our VAR model to assess the role of UMP on wealth inequalities in Great Britain.

We observe a significant jump in the Gini of financial net wealth following the announcement of QE1, while total net wealth and net housing wealth take accelerate after the announcement of QE2. Augmented net wealth presents a falling tendency up until the announcement of QE2 and then it raises.

\begin{table}[htbp]\centering
\def\sym#1{\ifmmode^{#1}\else\(^{#1}\)\fi}
\caption{Descriptive statistics of wealth variables (Mean values and percentiles)}
\def\arraystretch{1.3}%  1 is the default, change whatever you need
\label{tab:wealth_desc}
\begin{tabular}{l*{6}{c}}
\toprule
                    &     2006-07&     2008-09&     2010-11&     2012-13&     2014-15&        2016\\
\midrule
Financial Assets (formal)&       42297&       44529&       48657&       59433&       58198&       68055\\
Financial Assets (all)&       48360&       47884&       51553&       61849&       60343&       70570\\
Housing Assets (HMR)&      159878&      154519&      158940&      161879&      177993&      204163\\
Housing Assets (all)&      151421&      138971&      143503&      147523&      167079&      199300\\
Financial Liabilities&        3414&        3826&        4251&        3987&        4297&        4849\\
Housing Liabilities &       41540&       40860&       41364&       40646&       42334&       46495\\
Total Liabilities   &       44953&       44686&       45615&       44634&       46630&       51344\\
Total wealth (gross)&      202175&      199049&      207597&      221312&      236186&      272218\\
Net property wealth &      151421&      138971&      143503&      147523&      167079&      199300\\
Net financial wealth&       44280&       43661&       46830&       57315&       55493&       64945\\
Net wealth (augmented*) &      386670&      408941&      389569&      418850&      465097&      547112\\
Net wealth          &      195691&      182537&      190312&      204810&      222520&      264216\\
\midrule
Net wealth percentiles	&		&		&		&		&		&		\\
\midrule
10	&	-599	&	-1000	&	-850	&	-831	&	-850	&	-1400	\\
25	&	3500	&	3152	&	3000	&	2300	&	2500	&	3380	\\
50	&	113300	&	106367	&	104256.5	&	100338	&	106451	&	124780	\\
75	&	248406	&	237014	&	242287	&	251000	&	276020	&	327740	\\
90	&	450308	&	428750	&	444976	&	473000	&	526200	&	643000	\\
\midrule
Net housing wealth percentiles	&		&		&		&		&		&		\\
\midrule
10	&	0	&	0	&	0	&	0	&	0	&	0	\\
25	&	0	&	0	&	0	&	0	&	0	&	0	\\
50	&	97000	&	90000	&	90000	&	85000	&	90000	&	105000	\\
75	&	200000	&	189999	&	195000	&	200000	&	220000	&	259999	\\
90	&	341000	&	318998	&	340000	&	350000	&	400000	&	498500	\\
\midrule
Net financial wealth percentiles	&		&		&		&		&		&	\\	
\midrule
10	&	-4434	&	-6015	&	-6551.3&	-5748	&	-5928	&	-7391	\\
25	&	1	&	0	&	0	&	4	&	5	&	6	\\
50	&	5432	&	6154	&	5900	&	5550	&	5890	&	7390	\\
75	&	38050	&	38500	&	38501	&	39500	&	41110	&	52000	\\
90	&	113100	&	115890	&	115500	&	124089	&	133710.5	&	173400	\\
\midrule
Observations        &       14088&       19181&       21391&       20427&       19184&        4464\\
\bottomrule
\end{tabular}
	\begin{tablenotes}
	All nominal values are estimated in Great British Pounds. Population sampling weights have been applied.
	Augmented net wealth includes pension wealth, as well as the self-reported value informal financial assets and liabilities to our net wealth concept.  
\end{tablenotes}
\end{table}

\begin{landscape}
\begin{table}[htbp]\centering
\def\sym#1{\ifmmode^{#1}\else\(^{#1}\)\fi}
\caption{Inequality changes in Great Britain (2006-2016)}
	\label{tab:ineq_desc}
\def\arraystretch{1.2}%  1 is the default, change whatever you need
\begin{tabular}{l*{7}{c}}
\toprule
	&	2006-07	&	2008-09	&	2010-11	&	2012-13	&	2014-15	&	2016	&	\% Change 2006-2016	\\
	\toprule
Gini Indeces	&				&		&		&		&		&	\\	
\midrule
Net wealth	&	0.66	&	0.66	&	0.67	&	0.69	&	0.69	&	0.68	&	4.29\%	\\
Net financial wealth	&	0.89	&	0.89	&	0.91	&	0.91	&	0.91	&	0.91	&	0.49\%	\\
Net property wealth	&	0.63	&	0.63	&	0.64	&	0.65	&	0.66	&	0.66	&	4.34\%	\\
\midrule
Net wealth percentile shares	&		&		&		&		&		&		&		\\
\midrule
0-25	&	-0.62	&	-0.76	&	-0.67	&	-0.53	&	-0.54	&	-0.51	&	-16.99\%	\\
25-50	&	6.90	&	6.66	&	6.19	&	5.19	&	5.10	&	5.34	&	-22.59\%	\\
50-75	&	22.40	&	22.83	&	22.01	&	20.53	&	20.52	&	20.32	&	-9.28\%	\\
75-100	&	71.32	&	71.27	&	72.47	&	74.81	&	74.92	&	74.85	&	4.95\%	\\
\midrule
Net housing wealth percentile shares	&		&		&		&		&		&		&		\\
\midrule
0-25	&	-0.36	&	-0.40	&	-0.25	&	-0.16	&	-0.15	&	-0.10	&	-71.68\%	\\
25-50	&	7.15	&	6.83	&	6.42	&	5.40	&	5.19	&	5.59	&	-21.88\%	\\
50-75	&	24.36	&	24.83	&	24.21	&	23.49	&	22.36	&	21.73	&	-10.78\%	\\
75-100	&	68.85	&	68.73	&	69.62	&	71.27	&	72.59	&	72.78	&	5.7\%	\\
\midrule
Net financial wealth percentile shares	&		&		&		&		&		&		&		\\
\midrule
0-25	&	-4.26	&	-4.95	&	-5.19	&	-3.72	&	-4.28	&	-4.05	&	-4.99\%	\\
25-50	&	0.96	&	1.13	&	1.00	&	0.78	&	0.84	&	0.90	&	-6.17\%	\\
50-75	&	10.03	&	10.49	&	9.68	&	7.78	&	8.59	&	8.81	&	-12.19\%	\\
75-100	&	93.27	&	93.33	&	94.51	&	95.17	&	94.85	&	94.34	&	1.15\%	\\
\midrule
Top 10 \% share	&		&		&		&		&		&		&		\\
\midrule
Net wealth	&	46.11	&	45.44	&	46.83	&	49.95	&	49.37	&	48.76	&	5.76\%	\\
Net housing wealth	&	43.52	&	42.53	&	43.37	&	44.96	&	46.44	&	46.56	&	6.97\%	\\
Net financial wealth	&	70.81	&	70.19	&	72.62	&	76.49	&	73.76	&	72.39	&	2.23\%	\\

\midrule
Observations        &       14088&       19181&       21391&       20427&       19184&        4464\\
\bottomrule
\end{tabular}
\end{table}

\end{landscape}

%%%%%%%%%%%%%%%%%%%%%%%%%%%%%%%%%%%%%%%%%%%%%%%%%%%%%%%%%%%%%%%%%%%%%
\section{Model set-up}
\label{sec:model}

We estimate the impact of UMP shocks on wealth inequality by applying the following structural VAR model:

\begin{equation}
\label{eq:1}
{y_t} = c + \mathop \sum \limits_{j = 1}^p {y_{t - j}}{B_j} + {\nu _t}
\end{equation}
where $y_t$ is the matrix of endogenous variables, $B_j$ is the coefficient matrix, $c$ is the vector of constant terms, and ${\nu _{\text{t}}} \sim {\text{N}}\left( {0,\Sigma } \right)$. The covariance matrix of the residuals, $\Sigma$ can be decomposed as $A_0 A_0^{'}=\Sigma$, with $A_0$ representing the contemporaneous impact of the structural shocks, $\epsilon_t$, where $\nu_t=A_0 \epsilon_t$. The matrix $y_t$ contains industrial production (IP), the consumer price index (CPI), the spread between the 10-year Government bond minus the 3-month rate, the nominal effective exchange rate (NEER), and the wealth inequality measure. To determine the impact of UMP on wealth inequality we use the short-term shadow rate as a measure for the stance of monetary policy. 

We estimate our model by adopting a Bayesian approach. We use a natural conjugate prior as described in \cite{blake2012applied} and \cite{koop2010bayesian} while estimation is performed by using a Gibbs sampling algorithm to approximate the posterior distribution of the model parameters. As discussed in \cite{koop2010bayesian} and \cite{koop2017bayesian}, structural analysis is benefited from Bayesian VARs as without prior information, it is hard to obtain precise estimates of so many coefficients and as a result, features such as impulse responses and variance decompositions tend to be imprecisely estimated. 

In order to describe the priors and the algorithm used to estimate the BVAR, consider the compact form of equation \ref{eq:1}: 

$ Y_t=X_t B+\nu_t    \text{	where	}  X=\{c_{i},Y_{it-1},...,Y_{it-p}\} ,\text{	or	}$

$ Y_t=(I_{N}\otimes \)X) B+ V    
\text{	where	} $y = vec\left( Y_{t} \right)$, $b = vec\left( B \right)$, and $V = vec\left( {{\nu _t}} \right)$.

The prior for the VAR coefficients $b$ is normal and given by $p\left( b \right) = \sim N \left( {{{\bar b}_0},\Sigma  \otimes \Xi } \right)$. For ${\bar b_0}$, a conventional Minnesota scheme will be typically adopted, setting values around 1 for own first lag coefficients, and 0 for cross variable and exogenous coefficients. When setting $\Xi $ we approach a Minnesota type of variance matrix by adopting the strategy followed in \cite{karlsson2013forecasting} according to which $\Xi $ is a diagonal matrix where the diagonal elements are defined as $\sigma _{{\alpha _{ij}}}^2 = \left( {\frac{1}{{\sigma _j^2}}} \right){\left( {\frac{{{\lambda _1}}}{{{l^{{\lambda _3}}}}}} \right)^2}$ for lag terms, and $\sigma_j$  are variances approximated by individual AR regressions estimated via OLS. For exogenous variables, the variance is defined as $\sigma _e^2 = {\left( {{\lambda _1}{\lambda _4}} \right)^2}$. We set $\lambda_1 =0.1, \lambda_3 =1, \lambda_4 = 10^3$ that is typical in the literature. The posterior distribution of the VAR coefficients conditional on $\Sigma$ is normal, given by:

\begin{equation}
\label{eq:xxQ}
H\left( {b\backslash \Sigma ,{Y_t}} \right)\sim N\left( {{\mu ^*},{\nu^*}} \right)
\end{equation}

where:

${\mu ^*} = {\left( {{H^{ - 1}} + {\Sigma ^{ - 1}} \otimes X_t'{X_t}} \right)^{ - 1}}\left( {{H^{ - 1}}{{\bar b}_0} + {\Sigma ^{ - 1}} \otimes X_t'{X_t}\hat b} \right)$

and

${n^*} = {\left( {{H^{ - 1}} + {\Sigma ^{ - 1}} \otimes X_t'{X_t}} \right)^{ - 1}}$

where $\hat b$ denotes the OLS estimate of the VAR coefficients in vectorised format:  $\hat b = {\rm{vec}}\left( {{{\left( {X_t'{X_t}} \right)}^{ - 1}}\left( {X_t'{Y_t}} \right)} \right)$.

Next, regarding $\Sigma$, the conjugate prior is an inverse Wishart distribution with prior scale matrix $S$ and prior degrees of freedom $\alpha$, 
that is $p\left( \Omega  \right)\sim IW\left( S, \alpha  \right)$. Following \cite{karlsson2013forecasting}, we set $S = \left({a - n - 1} \right){\sigma_{d}}$ where ${\sigma_{d}}$ is a diagonal matrix with diagonal elements given by $\sigma_{\iota}^2$ and $\alpha=n+2$. Given this prior, the posterior for $\Sigma$ conditional on b is also inverse Wishart:

\begin{equation}
\label{eq:xx3}
H\left( {\Sigma \backslash b,{Y_t}} \right) \sim IW\left( {\Sigma ,{\rm T} + \alpha } \right)
\end{equation}

where $T$ is the sample size and

$\overline {\Sigma }  = \bar S + {\left( {{Y_t} - {X_t}B} \right)'}\left( {{Y_t} - {X_t}B} \right){\text{}}$

The Gibbs sampling algorithm consists of the following two steps. First, we sample the VAR coefficients from their conditional posterior distribution (Equation \ref{eq:xxQ}). Second, given $b$ from step 1, we draw $\Omega $ from its conditional distribution (Equation \ref{eq:xx3}). The algorithm is run for 100,000 iterations discarding the initial 60,000 to ensure convergence. Based on information criteria, we estimate the BVAR with four lags, considering alternative lag lengths in the robustness section.

Regarding the identification of the UMP shock, we follow the relative literature on the monetary policy transmission popularized by \cite{christiano1999monetary, christiano2005nominal} by adopting a recursive ordering of the variables based on the Cholesky decomposition of the $\Sigma$ matrix: $\Sigma = A_0 A_0^{'}$ as defined previously
ordering of the variables, the monetary policy rate is ordered after economic activity, inflation and the inequality index and before the yield spread and NEER. These restrictions on the macroeconomic variables are fairly standard in the literature and imply that output and prices react to monetary policy changes with a lag while a monetary policy shock is allowed to affect financial variables contemporaneously. Note that the results from the main specification are robust to different identification schemes that we describe later in the paper.

%%%%%%%%%%%%%%%%%%%%%%%%%%%%%%%%%%%%%%%%%%%%%%%%%%%%%%%%%%%%%%%%%%%%%%%%%%%%%%%%%%%%%%%%%%%%%%%%%%%%%%%%%

\section{Results}

\label{sec:results}

\subsection{The response of wealth inequality to UMP shocks}

\begin{figure}[!htbp]
	\caption{The impulse response of the Gini coefficient to a monetary policy shock. }
	\label{fig:giniinequality}
	\includegraphics[scale=.13]{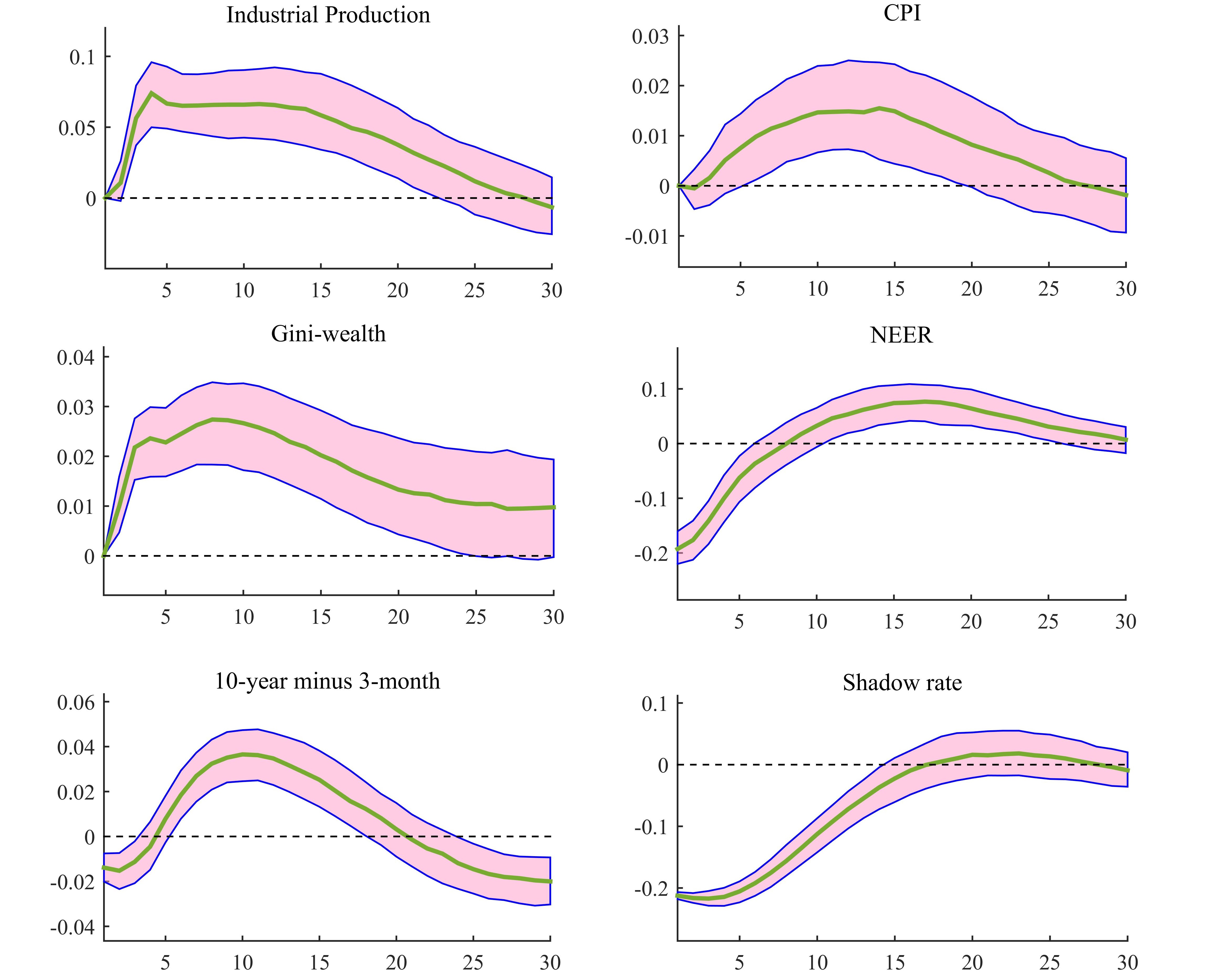}
		\begin{figurenotes}[Note]
		The vertical axis of each plot shows the responses in percent (apart from the shadow rate that is in percentage points). Time intervals on the x-axis are months. The green line is the median estimate and the pink shaded area depicts the 68 percent error bands. 
	\end{figurenotes}
\end{figure}

Figure \ref{fig:giniinequality} shows the impulse responses over 30 months of all variables to a 20 basis point cut in the shadow rate. We note that the shock leads to a significant increase of the net wealth Gini coefficient. The impact of UMP shocks on raising wealth inequality is long-lived given that the effect persists for the whole forecasting horizon. Regarding the magnitude of the effect, we observe that the UMP shock is estimated to increase the Gini coefficient by almost 0.03\% or 0.06 original units one year after the change in policy.

In terms of the responses of the core macroeconomic variables to UMP shocks, we note a positive reaction of IP by around 0.07\% four months after the shock and a gradual increase in the CPI that reaches its peak (increase of 0.015\%), one year after the shock. This result highlights the short-term benefits of UMP, that is, to support economic growth and boost inflation. Last, the 20 basis points reduction in the shadow rate elicits a fall in both the term spread and the nominal effective exchange rate on impact, as it would be expected. Similar responses have been obtained by \cite{hohberger2019distributional} after shocking an estimated DSGE of the EU economy with an expansionary monetary policy shock of both CMP and UMP. 

\subsection{Robustness checks of benchmark model}
\label{robustness_checks}

We test the robustness of the main findings by implementing an extensive sensitivity analysis. First, as estimated SSR series vary depending on modeling assumptions \citep{krippner2019}, we test the sensitivity of our estimates by employing an alternative SSR series as defined in \cite{krippner2014measuring} and \cite{claus2018asset}. Second, we re-estimate our model by providing alternative estimation of UMP shocks based on BoE total assets. A number of papers suggest that innovations on central bank assets do capture UMP innovations well during the period of the GFC \citep{gambacorta2014effectiveness, saiki2014does}. Although total assets capture only the quantitative effects of UMP, they remain observable variables and as such they serve as a good robustness check for our purposes. To properly identify the monetary policy shock, we adopt the identification scheme based on sign restrictions and described in \cite{baumeistera2013unconventional} and \cite{boeckx2017effectiveness}. Accordingly, we assume that the contemporaneous impact, of a shock that increases the balance sheet, on output and prices is positive while the same expansionary balance sheet shock decreases the yield spread and leaves the monetary policy rate unchanged. Regarding the mechanics of the process, we follow \cite{rubio2008structural} by letting ${\Sigma _t} = PDP'$  be the eigenvalue-eigenvector decomposition of ${\Sigma _t}$ and set ${\bar \Sigma _{0}}= P{D^{1/2}}$. We then draw an $r x r$ matrix $K$ from $N(0,1)$ and we compute $Q$ such that $K=QR$. Having these in hand, we compute the structural impact matrix as ${\Sigma _0} = {\bar \Sigma _0}Q'$ and retain it, if it satisfies the sign restrictions. Note that we also address the problem described in \cite{fry2011sign} under which the sign restrictions methodology presents impulse responses from different models rather than a single model and this could be misleading as a description of a single economy. To avoid this, we include a median target approach that selects the $A_{0}$ matrix that is closest to the median from a given number of draws from the algorithm. 
 
 Next, we replicate the benchmark analysis by using two alternative identification schemes. Specifically, we estimate a version of the model by adopting a stronger restriction that forces all the variables in the system to respond to UMP shocks with a lag, i.e. a cut in the shadow rate has a zero impact on all variables contemporaneously; we achieve this by ordering the shadow rate last and after the financial variables. Secondly, we use sign restrictions instead of recursive decomposition by assuming that an expansionary UMP shock leads to a contemporaneous decrease in the shadow rate and the yield spread, a rise in inflation and IP, and a decrease in the effective exchange rate (see also \cite{uhlig2005effects}).

Furthermore, we estimate two additional versions of the benchmark model, this time by checking whether our results are sensitive to the use of alternative inequality measures. We first re-estimate the benchmark model by replacing the Gini coefficient with the 20-20 ratio of net wealth. The 20:20 ratio compares the share of wealth of the 20\% wealthiest households with the share of wealth of the 20\% poorest, effectively depicting the wealth of the rich as a multiple of the poor’s wealth. In other words, the the 20:20 ratio is equal to $(100-Q_{80})/Q_{20}$, where $Q_{20}$ and $Q_{80}$ are the quantiles of order 0.2 and 0.8, respectively. In the second specification we use a popular alternative to the Gini, namely the coefficient of variation. The coefficient of variation is estimated by dividing the standard deviation of the net wealth distribution by its mean. The more equal a wealth distribution is, the smaller the standard deviation and consequently the coefficient of variation will be smaller in more equal distributions\footnote{According to \cite{cowell2017} the coefficient of variation is the only inequality measure within the Generalized-Entropy class that is applicable to wealth distributions. Yet, because by construction, the measure is very sensitive to outliers in the two tails, it is more appropriate when the distribution approaches normality. To address this problem for this robustness check, we top-coded the richest and the poorest 1\% of our sample and then estimated the coefficient of variation for household net-wealth. Although, this is admittedly a crude approach to correct for extreme values at the two tails, potentially leading to underestimation of inequality, it is still widely used in the relevant literature (see, \cite{mumtaz2017impact, coibion2017innocent}) and can help reducing measurement bias of the coefficient of variation estimates \citep{colciago2019}}.

Last, the findings are robust to perturbations to the benchmark VAR specification such as the addition of extra lags and the inclusion of additional endogenous variables. Specifically, in the former case, we re-estimate our model by including six lags. In the latter case, we augment the vector of endogenous variables by including a large number of macroeconomic and financial variables.\footnote{We augment the initial vector of endogenous variables by adding: the UK all share index, the crude oil price in dollars, imports, exports, government consumption, unemployment rate, weekly earnings, the 3-month Treasury bill rate, the 5-year Government bond yield and last, the 10-year and the 20-year Government bond yields} Note that now the size of the BVAR increases significantly. This inevitably leads to the curse of dimensionality that refers to the large number of parameters that have to be estimated in the model. We deal with this issue by adopting a dummy observation prior as described in \cite{bandura2010} and \cite{blake2012applied} to achieve Bayesian shrinkage. Practically, the hyperparameter that controls the overall tightness $\lambda$ of the prior distribution is set in relation to the size of the BVAR so that the higher the number of variables is, the more the parameters should be shrunk to avoid overfitting.

\begin{figure}[!htb]
	\caption{Robustness checks of the baseline result}
	\label{fig:robustness}
	\includegraphics[scale=.145]{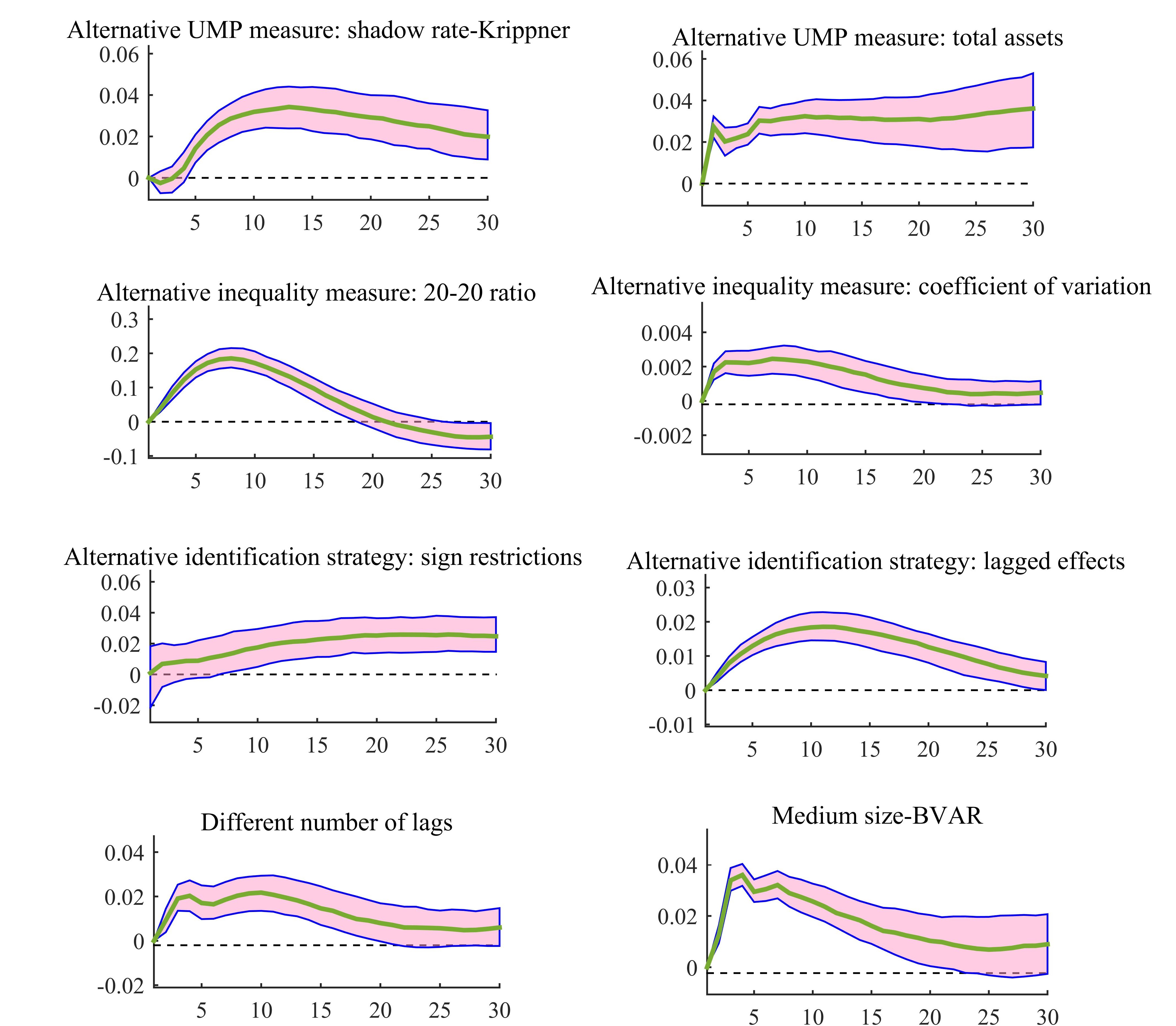}
	\begin{figurenotes}[Note]
The vertical axis of each plot shows the response in percent. Time intervals on the x-axis are months. The green line is the median estimate and the pink shaded area depicts the 68 percent error bands. 
	\end{figurenotes}
\end{figure}

 The results from these eight different specifications described above are depicted in Figure \ref{fig:robustness}. Note that we only show the response of the inequality measure as this is the variable of main interest. We observe that the responses of the inequality measure in all cases convey a similar message to the benchmark case, generating a significant increase of wealth inequality that persists over the forecasting horizon. Regarding the other variables of the system, the macroeconomic responses (not reported here but available upon request) are similar to the benchmark case with the UMP shocks generating a positive reaction of IP and CPI and a negative response of the term spread and the NEER on impact.
 
\newpage

\subsection{Heterogeneity of responses-Percentile groups}

\begin{figure}
	\caption{Impulse responses of net wealth quantiles}
		\label{fig:quantilie_shares}
	\includegraphics[scale=.6]{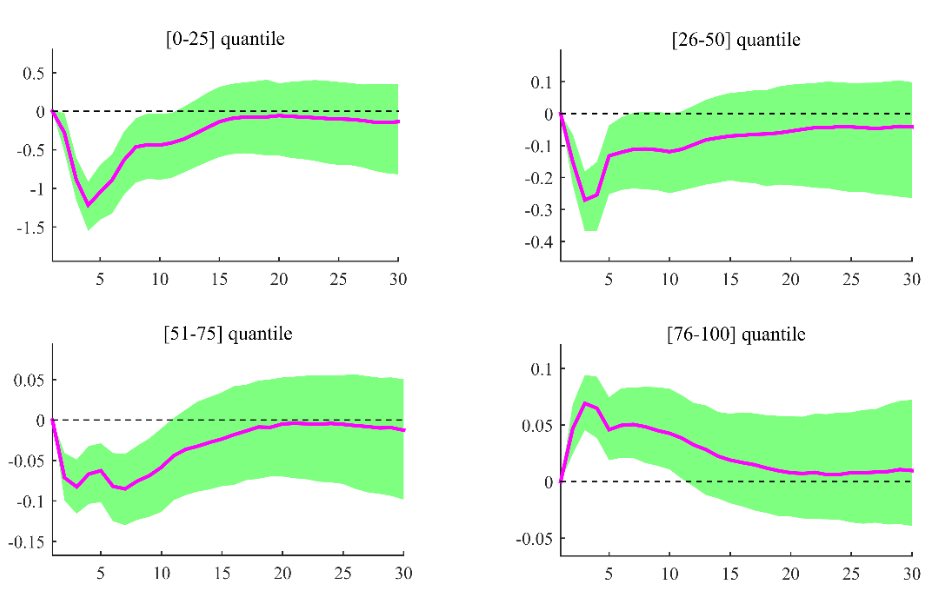} 
			\begin{figurenotes}[Note]
		The vertical axis of each plot shows the response in percent. Time intervals on the x-axis are months. The pink line is the median estimate and the green shaded area depicts the 68 percent error bands. 
	\end{figurenotes}
\end{figure}

% I removed the following sentence as I think we need to introduce it in a different way.
%Similar conclusions can be drawn by repeating the same exercise using household's net wealth quantile shares as an inequality measure.
In order to uncover potential reasons behind the response of wealth inequality on UMP shocks witnessed above, we examine how households at different points on the distribution respond to the same UMP shock defined above. The impulse responses of each share (i.e., 0-25; 26-50; 51-75; 76-100) to UMP shocks are shown in Figure \ref{fig:quantilie_shares}. Having in mind a model set-up with multiple agents across the asset and debt distribution similar to the one described in section \ref{sec:theory}, increases (declines) in the asset-rich portfolio values correspond to increasing (declining) wealth inequality. UMP shocks boost the value of assets and in turn the net wealth position of asset rich households. This is translated in our empirical model to increases in the net wealth Gini (Figure \ref{fig:giniinequality}), which, as can be seen by observing the bottom right panel of Figure \ref{fig:quantilie_shares}, is due to increases to the quantile share of the upper households in the net wealth distribution, i.e. those standing on the fourth quantile (76-100). As can be observed in figures \ref{fig:financial_shares} and \ref{fig:housing_shares} in Appendix \ref{sec:additional}, similar patterns emerge if we decompose the wealth shares in their two main components, financial and housing net wealth. According to figure \ref{fig:financial_shares}, the asset price inflation effect for the richest quantile is positive and significant in the first year after the shock. Following the theoretical channels described earlier, a possible explanation for this result is that financial asset owners that have originally switched to short-term instruments when re-balancing their portfolios, effectively lose coupon payments from long-term bond holdings. It is also crucial to stress that although the effect on the top quantile is falling, it never gets negative as in the poorer quantiles. This comes in contrast to theoretically estimated studies and suggests that the revaluation effect is stronger than the capital income losses effect \citep{hohberger2019distributional}. 

%Figure \ref{fig:ginicomponents} presents our baseline results, decomposed to the Gini coefficients of the two main components of net wealth, namely financial and property wealth.
%
%
%\begin{figure}
%	\hfill
%	\subfigure[Net Financial Wealth]{\includegraphics[scale =.45]{2_1_component_financial.png}}
%	\hfill
%	\subfigure[Net Property Wealth]{\includegraphics[scale=.45]{2_2_component_property.png}}
%	\hfill
%	\caption{The effects of UMP on the components of inequality}
%	\label{fig:ginicomponents}
%	\begin{figurenotes}[Source]
%		Authors' estimations based on WAS Data (ONS, 2018).
%	\end{figurenotes}
%	
%\end{figure}

%Wouldn't one expect to have at least one component (the financia) to have a higher response than the overall gini?

\subsection{Unconventional monetary policy on Wealth inequalities: variance decomposition}

\begin{figure}[!htbp]
	\caption{Percentage contribution of unconventional monetary policy shocks to the forecast error variance of the net wealth Gini coefficient.}
	\label{fig:var_dec}
	\includegraphics[scale=.45]{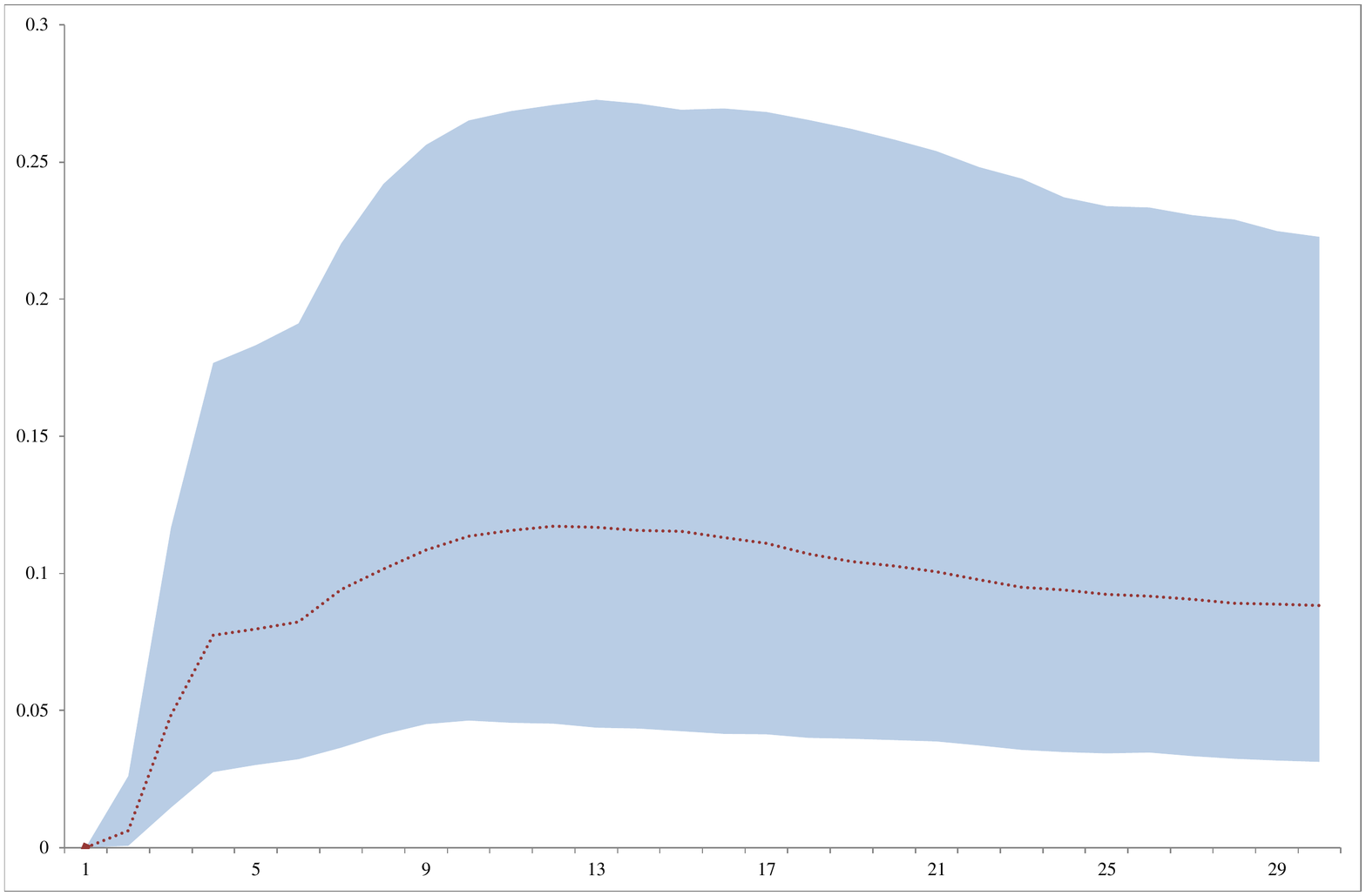}
	\begin{figurenotes}[Note]
	Percentage contribution of UMP shocks to the forecast error variance of the Gini coefficient.The red dotted line is the median estimate and the light blue shaded area depicts the 68\% error
	band.

	\end{figurenotes}	
\end{figure}

Another way to highlight the role played by UMP shocks in driving fluctuations in the Gini coefficient is by looking at the forecast error variance decomposition. Figure \ref{fig:var_dec} plots the contribution of the UMP shock to the forecast error variance of the inequality measure. The red dotted line shows the median estimate and the light blue shade area is the 68\% error bands. We observe that the median contribution of the UMP shock to fluctuations in the wealth Gini is around 11\% at the first year horizon, suggesting that UMP measures did play an important role in the widening of the wealth gap over the 2009-2016 period.

%%%%%%%%%%%%%%%%%%%%%%%%%%%%%%%%%%%%%%%%%%%%%%%%%%%%%%%%%%%%%%%%%%%%%%%%%

\subsection{Unconventional monetary policy on Wealth inequalities: capturing the channels}

Next, we augment our main BVAR specification by adding five variables in order to examine the transmission channels through which UMP affects wealth inequality. First, as discussed in Section \ref{sec:theory}, the impact of central bank asset purchases on equity prices and returns is established in the UMP literature and functions through the portfolio-rebalancing channel \citep{joyce2012, mclaren2014using, neely2015unconventional}. The announcement of central bank purchases implies a simultaneous increase in asset prices and a drop in corporate bond yields \citep{joyce2012, neely2015unconventional}. On the one hand, government bond yields lead to lower corporate bond yields for a given corporate bond spread to compensate for the risks of holding corporate bonds relative to government bonds. On the other hand, government bond yields increase the present value of future dividends, thus raising equity prices. Furthermore, as investors rebalance their portfolios from risk-free government bonds to more risky assets, reduces the risk-premium for holding equities, and therefore, puts upward pressure on their prices \cite{joyce2011financial, joyce2012}. To capture the simultaneous impact of QE on corporate bond yields and equity prices, we use the "UK all share index" taken from the FRED database and the Sterling (GBP) corporate bond yields on all issuers (including financials) rated AAA-BBB that, taken from the millennium of macroeconomic database of BoE, respectively.\footnote{See, \url{www.bankofengland.co.uk/publications/quarterlybulletin/threecenturiesofdata.xls}} Both variables allow us to quantify the impact of the portfolio rebalancing channel on wealth inequality changes.

%housing channel
Second, we investigate the channel that works via effects of higher asset prices through housing wealth. To do this, we augment our BVAR with the housing market index taken from Gov.UK as an additional variable. As mentioned earlier, UMP can affect the housing market by inflating house prices and boosting residential investment \citep{rahal2016housing}. Given the more even distribution of housing wealth compared to financial wealth, the expected effect of UMP on housing wealth inequality can go either direction. Third, we look at the savings redistribution channel as proxied by the effect of a decreased borrowing rate on the net wealth position of borrowers and savers \citep{inui2017effects, CASIRAGHI2018215}. Our hypothesis is that when the savings redistribution channel is open, the QE impacts on the borrowing rate and borrowers are likely to be better off as interest payments on debts fall more than interest payments on savers' deposits.  We use the mortgage interest rate and the unsecured loan rate taken from the millenium macroeconomic database of BoE in order to uncover the potential impact of this channel. 

%consider to create a new net interest income variable that is the difference between the borrowing rate and the deposits rate in order to capture both effects on borrowers and savers adequately. 

We build the counterfactuals as hypothetical impulse responses which depict only the direct impact of UMP shocks on inequality and neutralize the indirect effect through each of the three transmission channels. This is done by constructing a counterfactual sequence of shocks to the variables such that the impulse response of each of the additional variables described above, to UMP shocks, is equal to zero at all horizons.  The comparison of the counterfactual responses of Gini with the actual responses estimated in the unrestricted model give us a statistical measure of the importance of each of these channels on the transmission of UMP to households' wealth.  

\begin{figure}[!htb]
	\caption{Unconventional monetary policy on wealth inequality: transmission channels}
	\label{fig_channels}
	\includegraphics[width=1\textwidth]{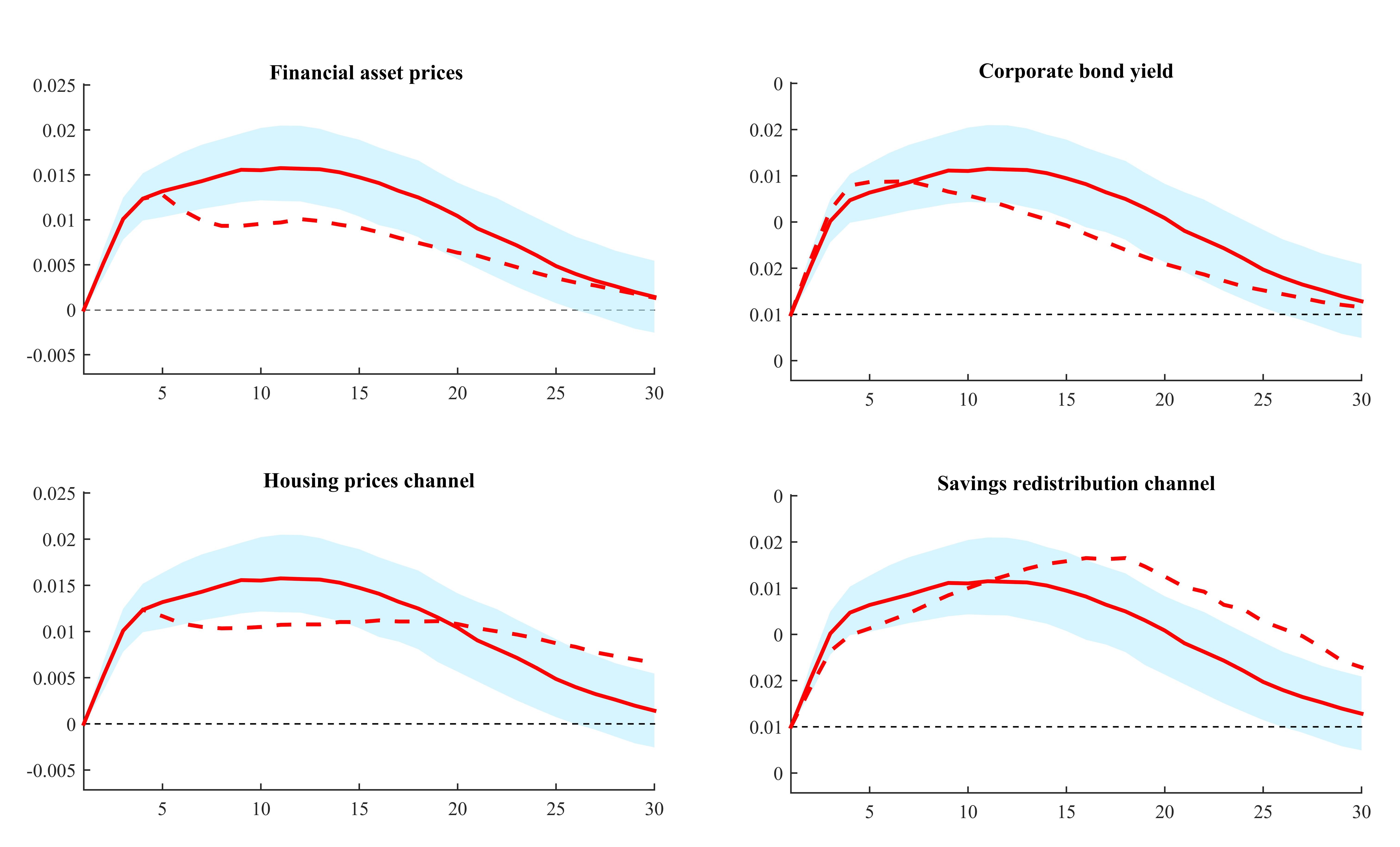} 
\begin{figurenotes}[Note]
	The red line show the median estimate and the blue shaded area is the 68\% error
	band. The dotted lines show the median response of Gini from the counterfactual experiment.
\end{figurenotes}	
\end{figure}

Figure \ref{fig_channels} show the results. The red solid lines represent the median responses of Gini computed in the unrestricted BVAR together with the 68\% confidence bands while the dotted lines show the median response of Gini from the counterfactual experiment. We observe that shutting down the share index and the corporate spread to UMP shocks lead to significant declines of the responses of Gini (note that both counterfactual responses are below the responses of the unrestricted model while they also lie outside the error bands for most of the forecasting horizon), indicating that the portfolio rebalancing channel has an important role in increasing wealth inequality. Our results confirm empirical works suggesting a positive effect of the portfolio composition channel on wealth inequality {\cite{domanski2016wealth, CASIRAGHI2018215, adam2015price, lenza2018does} for other countries, but contradict theoretical predictions implying that the effect is short-lived and changes direction in the medium-run \citep{hohberger2019distributional}. 

Next, we observe that when the housing prices channel is shut down, the counterfactual response of Gini in the first twenty months after the shock is lower in magnitude compared to its counterparts in the unrestricted case, thus exerting extra pressure on the widening of the wealth gap. This finding comes in contrast to most of the simulations literature that predict housing price increases offsetting the regressive outcomes of financial asset inflation \citep{Adam2016, lenza2018does, bivens2015, pugh2018distributional}. As this finding is partly driven by the share of households in Great Britain with no housing assets at all (around 25\% according to our estimates), it adds to the policy relevant literature suggesting that the promotion of home-ownership for lower wealth groups should lead to lower wealth inequality (see, \cite{kaas2019does}). %Note also that two years after the shock, the pattern is reversed and the counterfactual response of Gini is above the unrestricted case. However, the effect is marginally statistically significant only toward the end of the forecasting horizon. 
 The reverse effect is observed when we switch off the savings redistribution channel (i.e. setting the coefficients of both the mortgage interest rate and the unsecured loan rate, to zero). The counterfactual response of Gini is significantly higher in the first year following the shock, indicating that the savings redistribution channel acts as a counterbalancing force since the fall in wealth inequality offsets the upward pressures on inequality elicited by the other two channels.

\subsection{Counterfactual policy analysis}
\label{sec:counterfactual}

The analysis so far has focused on the role of UMP shocks in driving wealth inequality in the zero lower bound period. In this section, we explicitly measure the impact of quantitative easing (QE) on inequality by focusing on the period before the implementation of the last round of quantitative easing in August 2016, where the BoE had already completed a total of GBP 435 billions of purchases. In particular, we are interested in exploring what would have happened to the wealth inequality if the BoE had reversed its QE policy earlier. In order to measure the impact of QE on the Gini coefficient, we use our benchmark BVAR model to simulate the economy one period ahead conditional on specific counterfactual policy paths. Specifically, the BVAR model is estimated from 2009:01 up to 2014:12 and then it is used to carry out two counterfactual experiments from 2015:01 to 2016:06. Each counterfactual experiment involves two conditional forecasts that we will call them the 'QE-scenario' and the 'non-QE scenario'. 

In relation to the first counterfactual experiment, we replace the shadow rate with the policy rate in the benchmark BVAR in 5.1, and, following \cite{kapetanios2012assessing} we assume that QE affects the economy by reducing the yield on long-term government bonds. Therefore, the first conditional forecast assumes that the path of the term spread is higher than observed by 100 basis points over the forecasting horizon (non-QE scenario); this could be considered as a quantitative tightening policy. The second conditional forecast assumes that the term spread equal its observed value over the forecasting horizon (QE scenario). 

With respect to the second counterfactual experiment, we now assume that QE affects the economy by reducing the shadow rate. 
The first conditional forecast uses the shadow rate as a measure of monetary policy over the forecasting horizon (QE scenario) while in the second conditional forecast, we replace the values of the shadow rate with the actual policy rate over the forecasting horizon (non-QE scenario); this latter can be seen as a quantitative tightening policy.

\begin{figure}[!htb]
	\centering
	\begin{subfigure}
		\centering
		\caption{The impact of QE on Gini coefficient when the term spread is the measure of monetary policy }
		\label{fig:counter_1}
		\includegraphics[width=.8\textwidth, height = 6cm]{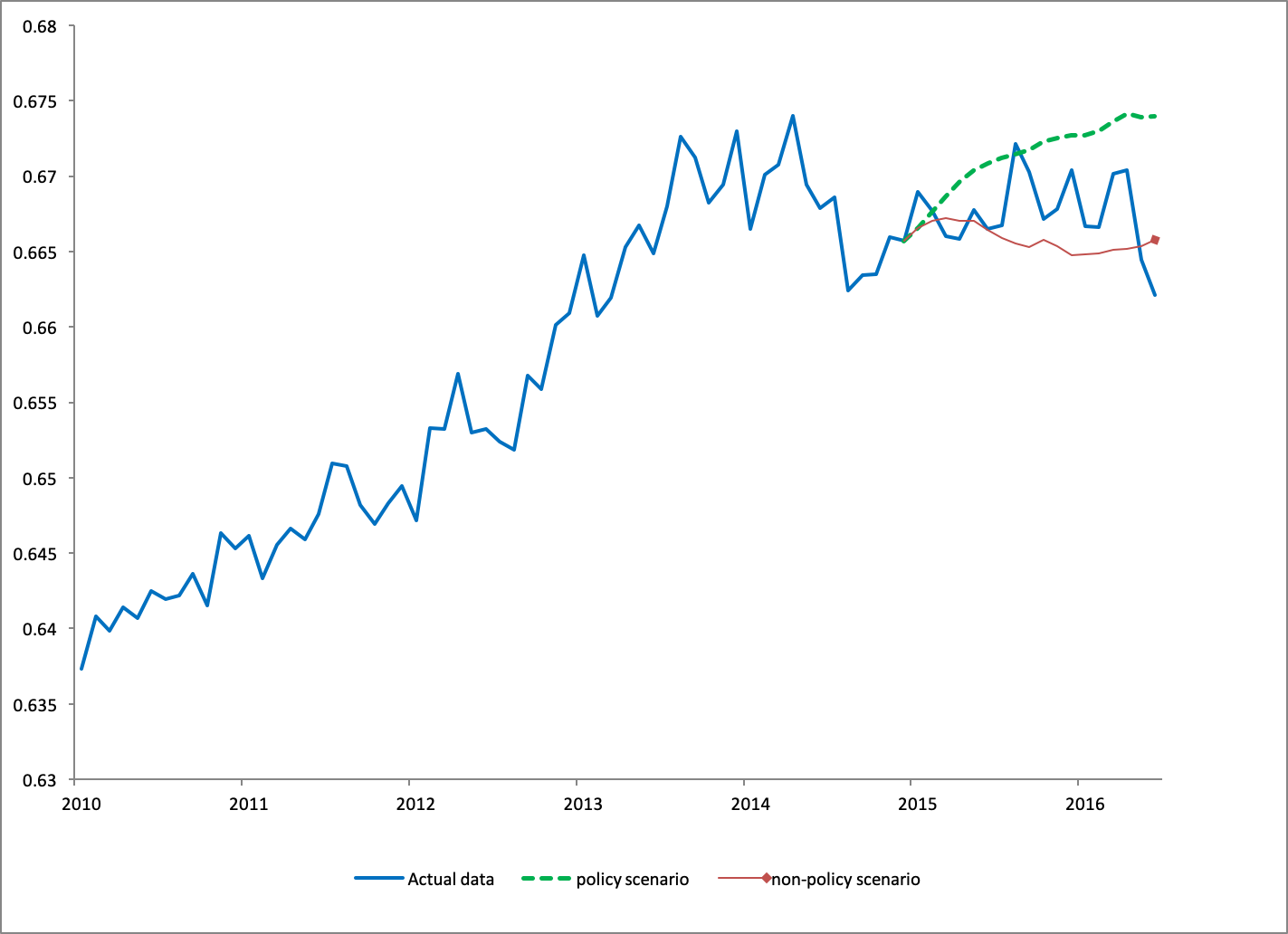}

	\end{subfigure}%
	\begin{subfigure}
		\centering
		\caption{The impact of QE on Gini coefficient when the shadow rate is the measure of monetary policy}
		\label{fig:counter_2}
		\includegraphics[width=.8\linewidth, , height = 6cm]{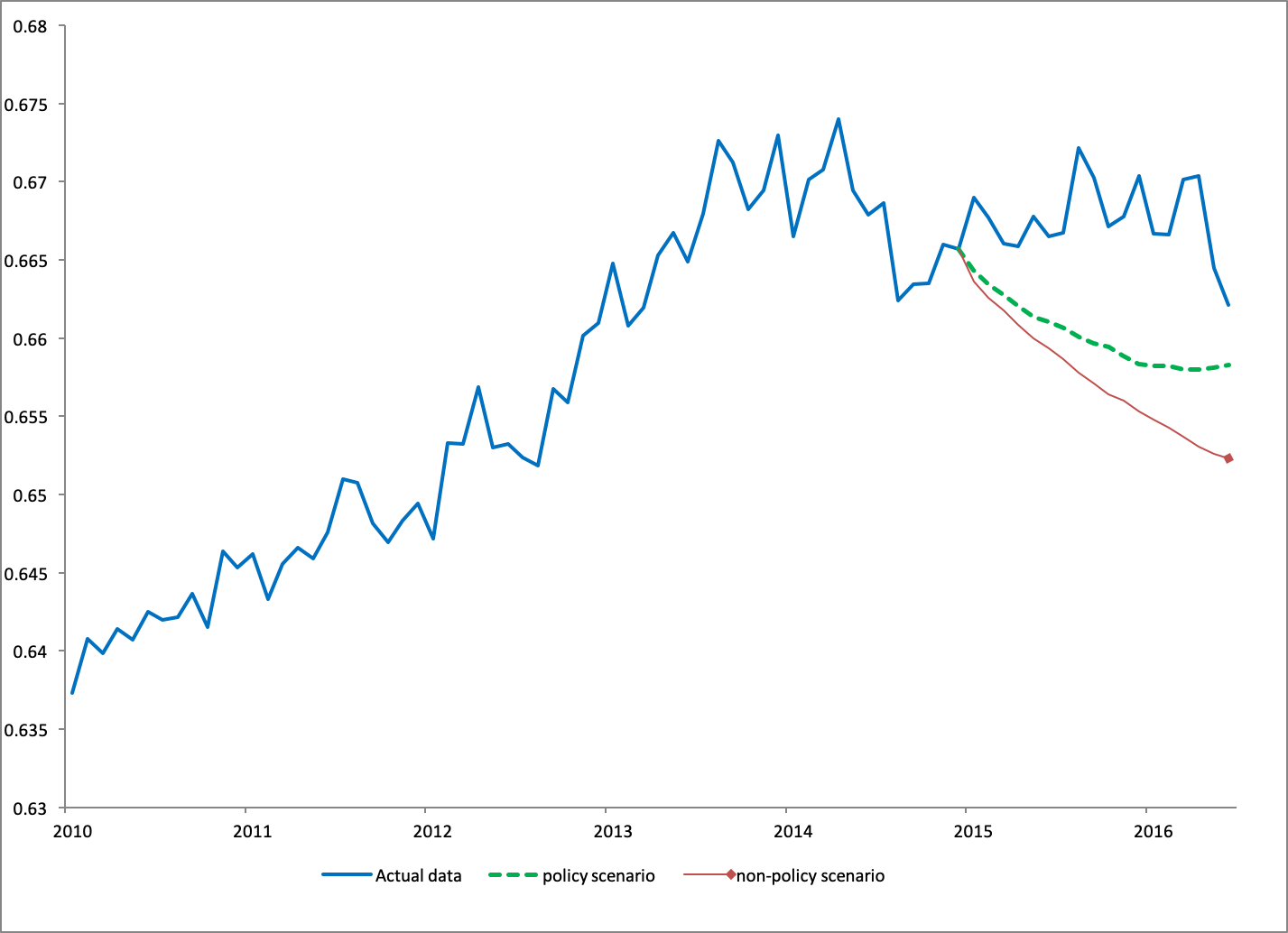}
	\end{subfigure}
		\begin{figurenotes}[Notes for Figures 8,9]
	The  blue  line  shows  the  actual  data  for  the  Gini coefficient, the red line shows the median forecast  under the non-QE scenario while the green dotted line shows its median forecast under the QE scenario.
		\end{figurenotes}
	\begin{subfigure}
		\centering
		\caption{The impact of QE on Gini coefficient from a TV-VAR model}
		\label{fig:counter_3}
		\includegraphics[width=\linewidth, , height = 6cm]{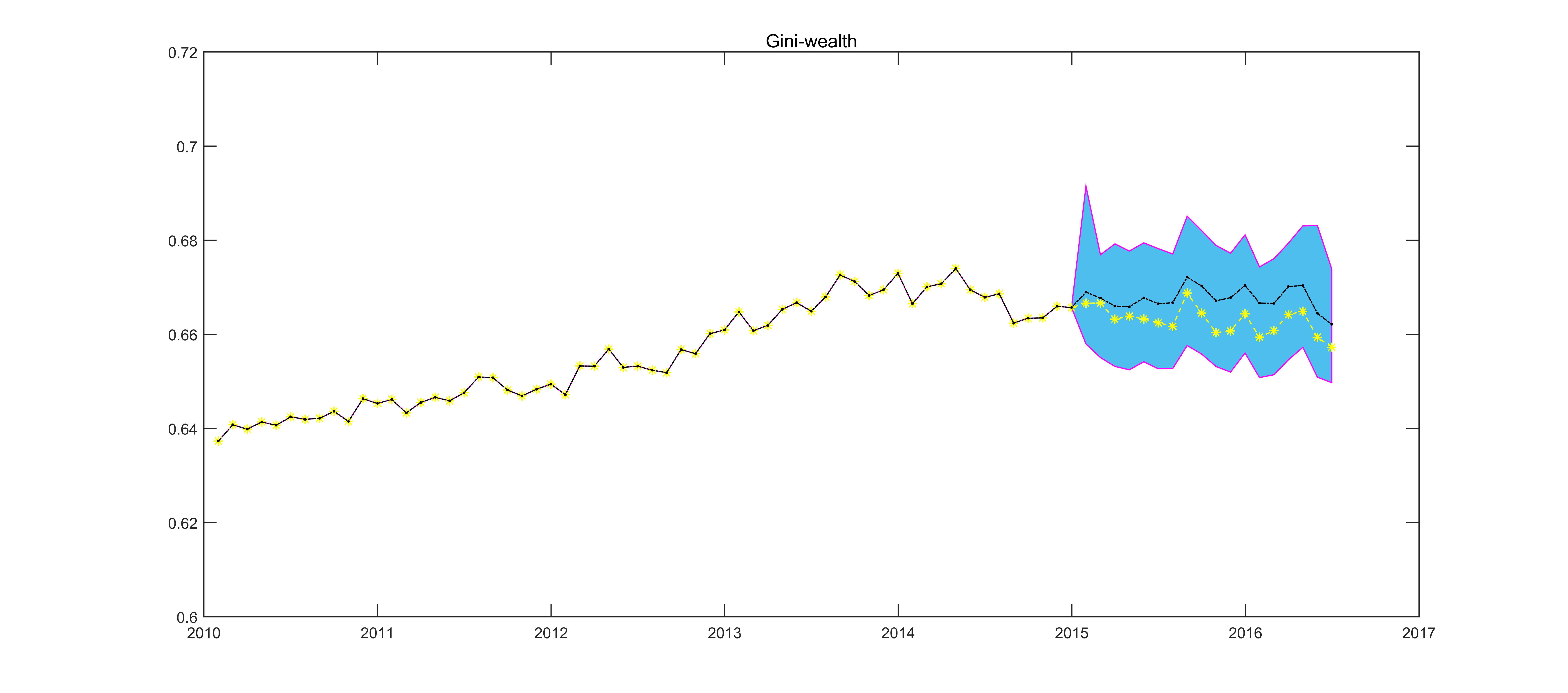}
	\end{subfigure}
		\begin{figurenotes}[Note]
The yellow starred line shows the counterfactual path while the blue line shows the actual data. The light blue shaded area depicts the 68\% bands.
	\end{figurenotes}
	\label{fig:counter_all}
\end{figure}

Figures \ref{fig:counter_1} and \ref{fig:counter_2} illustrate the results. The blue line shows the actual data for the Gini coefficient, the red line shows the median conditional forecast of the Gini coefficient under the non-QE scenario while the green dotted line shows its median conditional forecast under the QE scenario. We can see from both figures that the forecasts of the wealth Gini coefficients are much higher in the QE scenario as opposed to the non-QE scenario, throughout the forecasting period. The results suggest that UMP measures inflated the wealth of the rich and therefore widened inequality. 

We should highlight that this approach is based on out of sample forecasts from the BVAR model, thus the forecast distributions linked with both scenarios are extremely wide indicating that the estimates are uncertain. To check the robustness of our results, we adopt a different, structural approach based on  \cite{baumeistera2013unconventional}. Specifically, we employ a time varying VAR model (TV-VAR) by using the policy rate, the term spread, the Gini coefficient, inflation and IP. We describe the model in detail in the Appendix \ref{app:tvvar}. The model is estimated from 2009:01 to 2016:06. The innovation here with respect to the previous two approaches is that we use sign restrictions in order to be able to identify a shock to the term spread. In particular, following the identification scheme proposed by  \cite{baumeistera2013unconventional}, we assume that a contractionary monetary policy shock, consistent with quantitative tightening, increases the term spread, contemporaneously leads to a fall in inflation and IP but leaves the policy rate unchanged. At the same time, we assume that a conventional monetary policy shock raises the policy rate and leads to a fall in the term spread, inflation and IP contemporaneously. Having identified the term spread shock, we conduct a counterfactual experiment from 2015:01 to 2016:06 where we scale the shock such that the counterfactual value of the term spread is higher than the actual value by 100 basis points; in essence, what we get is the counterfactual path in the absence of QE.

Figure 	\ref{fig:counter_3} presents the results. Note that in order to examine whether QE could have affected the wealth Gini coefficient, we need to compare the implied path of the Gini coefficient obtained by the counterfactual scenario as described above with the actual path. The yellow starred line shows the counterfactual path while the blue  line shows the actual data. The light blue shaded area depicts the 68\% bands linked to the simulation. Our results suggest that during the whole simulation period the counterfactual distribution of the Gini coefficient is consistently below the actual data. This finding is in line with our results obtained from the previous approach, corroborating our hypothesis that UMP aggravated wealth inequality.

\subsection{Conventional Monetary Policy versus Unconventional Monetary Policy on wealth inequality}
\label{conv_unconv}

As discussed in section \ref{sec:theory}, the portfolio composition channel is the most crucial one for the transmission of unconventional monetary policy to the wealth holdings, and this takes place under periods through portfolio rebalancing by investors, which, in turn, affects the price of assets held by households. Still, the other main channel, namely savings redistribution via the borrowing rate, does function also under periods of CMP. The evidence from the literature is mixed when it comes to the distributional consequences of CMP, and most of the studies suggest that if any relationship exists, it favours the poorer parts for the wealth distribution \citep{meh2010aggregate, doepke2006inflation, adam2015price}, with only a few studies suggesting the opposite case (for example, \cite{BAGCHI201923}). 

To strengthen our case for the redistributional consequenses of UMP, we investigate whether the effects of monetary policy shocks on wealth inequality differ depending on two monetary policy states, i.e. the ZLB state where UMP measures are implemented and the non ZLB state where CMP is implemented. To do this, we use a Bayesian threshold VAR (TVAR) model that allows us to endogenously identify these two monetary policy states with respect to one transition variable that, in our case, is the shadow rate. The two different monetary policy states are determined by the value of this threshold variable with respect to a certain threshold that is estimated within the model. We use the six variables as defined in the benchmark specification while the monetary policy shock is identified using the same recursive identification scheme discussed in  Section \ref{sec:model}. The model is run over the full sample period, i.e. from 07/2006 to 06/2016. We describe the model in detail in the Appendix \ref{app:threshold_var}.

Figure 	\ref{fig:cmp_ump}  presents the impact of a 20 basis points decrease in the shadow rate on the Gini coefficient, in both conventional and unconventional monetary policy regimes. Before we discuss this figure, it is worth inspecting Figure \ref{fig:threshold}. The graph reveals that our TVAR successfully identifies the two monetary policy regimes that took place from the beginning of our sample until 2016, in the UK. In particular, the estimated threshold and the threshold variable show that regime one persisted up to early 2009 while from that point onward, regime two prevailed. Therefore, we interpret the negative shadow rate shock in regime one as representative of expansionary monetary policy via conventional measures, i.e. typical interest rate cuts, while we interpret the same shock in regime two as indicative of UMP shocks. 

\begin{figure}
	\caption{Estimated threshold against the threshold value}
		\label{fig:threshold}
	\includegraphics[scale=.8]{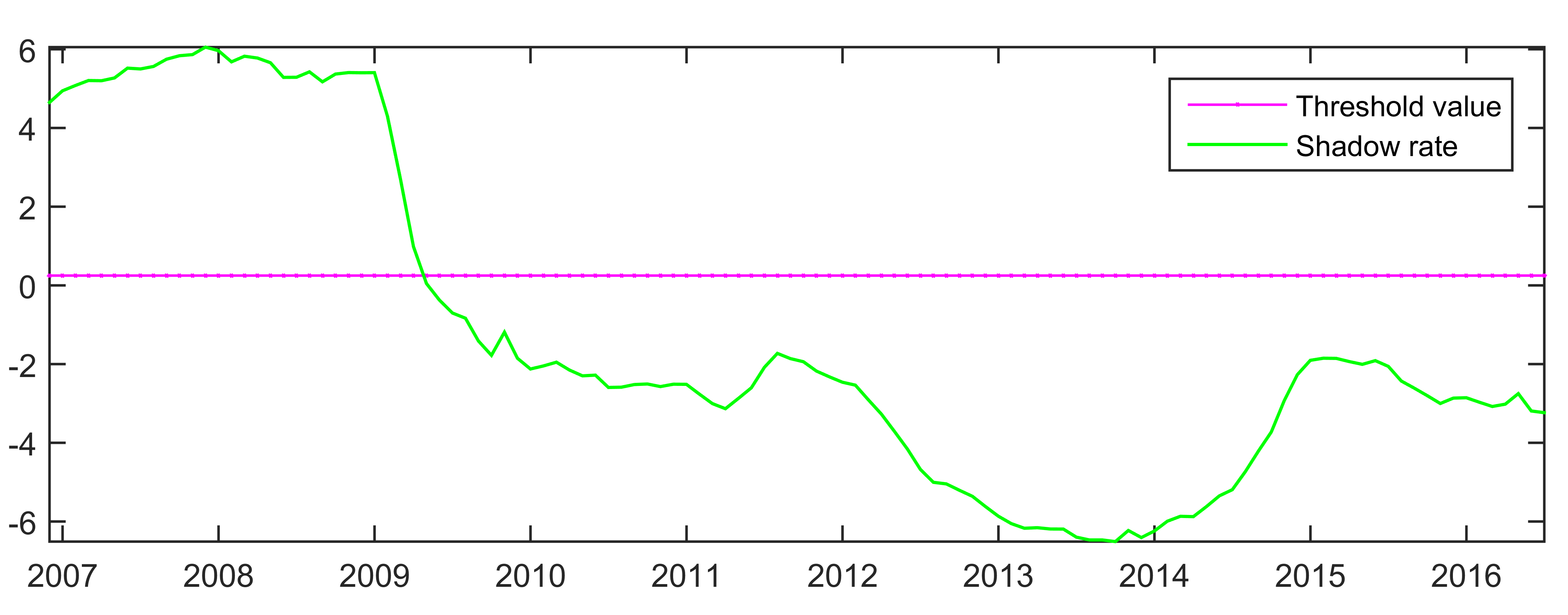} 
\end{figure}

\begin{figure}[!htbp]
	\caption{Impulse responses of CMP and UMP on the Gini coefficient of wealth inequality}
	\label{fig:cmp_ump}
	\hfill
	\subfigure[UMP]{\includegraphics[scale =.3]{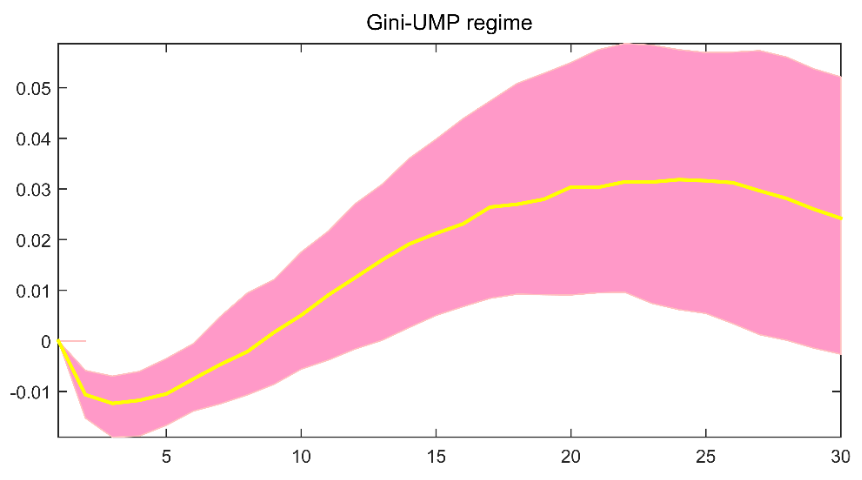}}
	\hfill
	\subfigure[CMP]{\includegraphics[scale=.3]{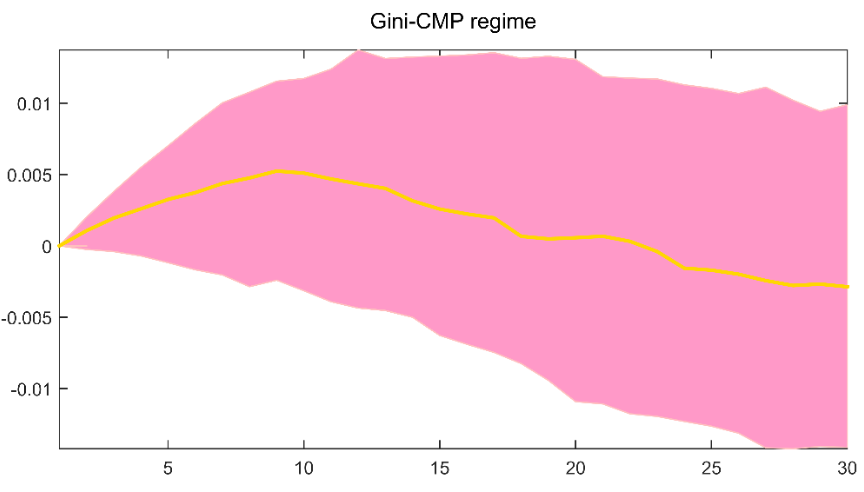}}
	\hfill
	\begin{figurenotes}[Note]
		The vertical axis of each plot shows the response of Gini coefficient in percent. Time intervals on the x-axis are months. The yellow line is the median estimate and the pink shaded area depicts the 68 percent error bands. 
	\end{figurenotes}
	%\caption{Wealth inequality in the UK (2006 - 2016)}
%	\begin{figurenotes}[Source]
	%	Authors' estimations based on WAS Data (ONS, 2018).
%	\end{figurenotes}
	
\end{figure}

Back to Figure 	\ref{fig:cmp_ump}, the right graph shows that in the conventional policy regime, an expansionary policy shock leads to a gradual rise in the Gini coefficient; note however that the null hypothesis that this effect equals zero cannot be rejected. On the other hand, the left graph shows that after a short-lived initial fall, we observe  a continuous increase of the Gini coefficient that becomes positive and significant, reaching approximately 0.03\% after two years (note that the estimated figure is very close to the one obtained under our benchmark model). Overall, the results indicate that conventional interest rate cuts up to early 2009 did not worsen wealth inequality while the alternative policy measures implemented since that date, have contributed to the surge of wealth inequalities.

Our results are in line with the evidence coming the existing literature suggesting that expansionary CMP does not increase wealth inequality. Yet, they should be read with caution as our dataset covers only a fraction of the long period during which CMP was implemented before the bank rate touched zero level. Thus, further work is needed that delves deeper to analyzing the effect of CMP on wealth inequality  from a historical perspective.\footnote{\cite{BAGCHI201923} and\cite{leroy2019redistributive} are recent works on these grounds.} Still, the juxtaposition of the two different monetary policy states using the same model, clearly strengthens the case put forward in this study in favour of a regressive redistributive effect of UMP

%%%%%%%%%%%%%%%%%%%%%%%%%%%%%%%%%%%%%%%%%%%%%%
\section{Concluding Remarks}
\label{sec:conclusion}
%%%%%%%%%%%%%%%%%%%%%%%%%%%%%%%%%%%%%%%%%%%%%%

We illustrated that UMP shocks have significant effects on wealth inequality: an expansionary monetary policy in the form of asset purchases raises the observed inequality across households as measured by their Gini coefficients and quantiles of net wealth, housing wealth, and financial wealth. Additional counterfactual policy experiments confirm the QE did play an important role in the widening of the wealth inequality gap. With respect to the pass through mechanism of UMP shocks, portfolio rebalancing is the most prominent channel in increasing wealth inequality and as our results show, it is activated through wealth effects via higher financial asset prices and drops in corporate bond yields.
%The pass-through of the MP shock on wealth inequality is far more prominent in the financial wealth component and is activated through the financial assets channel (FTSE). 

We also presented evidence of a regressive distributive effect activated through the housing prices channel. This result contradicts the intuition of theoretical and empirical works investigating the relationship between monetary policy and wealth inequality that typically predict housing revaluation effects offsetting the financial asset revaluation effects of UMP (see for example, \cite{Adam2016, lenza2018does,  pugh2018distributional}). Our results echo the relatively small share of home-owners in the real wealth distribution of Great Britain and the effect of UMP on house prices which takes more time to reveal \citep{rahal2016housing} compared to the revaluation effect of financial assets. In addition, we presented evidence in favour of the savings redistribution channel which transfers wealth from richer savers to poorer borrowers whose balance sheet items track variable interest rates.  Nevertheless, this channel is not strong enough to counterbalance the upward pressures on inequality driven by the inflationary pressures of the portfolio rebalancing channel and the housing price channel. Last, as expected, the results reveal that UMP is shown to present more regressive redistributive effects than CMP. This is most probably due to the existence of the portfolio rebalancing channel through large-scale asset purchases that is the major transmission channel under the zero lower bound.

 While our analysis captured a rich set of dynamics, it narrowed its focus to a partial equilibrium perspective, by evaluating the impact of UMP shocks on wealth distribution variables rather than focusing on their impact on other welfare measures such as income or consumption. These measures are most likely affected by UMP in a heterogeneous way, depending on the income structure and the consumption patterns of the economy. For instance, \citep{hohberger2019distributional} suggests the presence of feedback effects that potentially arise from monetary policy in general equilibrium, which may not be captured in an empirical analysis as in the case of a dynamic macroeconomic model. Yet, evidence from \cite{mumtaz2017impact} complements our analysis by providing evidence that QE worsened income and consumption inequality in the UK, on top of wealth inequalities addressed here. Taking all the evidence together, the message is that QE measures in the UK worsened overall economic inequalities.

Our results have important policy implications at a time when major central banks are moderately switching from unconventional to conventional monetary policy. Although UMP measures have proven to be a powerful monetary instrument to boost liquidity and investment when the zero lower bound is binding, they are accompanied with undesirable side effects, namely widening wealth disparities. Policies that induce regressive redistribution outcomes may not be desirable for the UK, given the central role of underlying social inequalities on the rise of populism \citep{goodwin2016, piketty2018brahmin} and possibly on the British referendum vote to leave the EU \citep{fetzer2018did}. 

Alternatives to UMP have been proposed, including fiscally oriented transfer policies by crediting (poorer) households with means-tested stipends, refundable tax credits targeted to poorer families and, unemployment insurance extensions.\footnote{See for example \cite{muellbauer201420} and \cite{baldwin2016helicopter} among others} Although the effectiveness of fiscal transfers in raising output and stabilizing investment under a binding ZLB is still debated \citep{eggertsson2012debt, mehrotra2018fiscal}, such measures are likely to have less regressive side-effects when it comes to wealth redistribution, due to the lower prevalence of the asset revaluation channel documented in the present study. Thus, the redistributive effects of fiscal versus monetary policy under the ZLB serve as a straight avenue for future research. The study of \cite{bivens2015} focusing on income inequality for the US case is a good starting point on this research agenda.

Finally, growing theoretical and empirical work suggests that wealth distribution disparities, on top of being an outcome of monetary policy, functions as a transmission mechanism of monetary policy to consumption through varying marginal propensities to consume for different net wealth profiles \citep{auclert2017monetary}. Thus, an avenue for future research would be the investigation of monetary policy-induced inequality as an endogenous variable in the monetary policy transmission and its effects on the economy-wide equilibrium.

%\openup -.5em

% Bibliography

\bibliographystyle{aer}
\bibliography{ineq_library}

\newpage
% The appendix command is issued once, prior to all appendices, if any.
\appendix
%%%%%%%%%%%%%%%%%%%%%%%%%%%%%%%%%%%%%%%%%%%%%%%%%%%%%%%%%%%%%%%%%%%%%%%%%%%%%

%
%
%\begin{figure}
%	\caption{Impulse responses on wealth quantiles: Financial net wealth}
%\includegraphics[scale=.6]{3_2_quant_financial_wealth.png} 
%\end{figure}
%
%
%
%
%\begin{figure}
%	\caption{Impulse responses on wealth quantiles: Housing net wealth}
%\includegraphics[scale=.4]{3_3_quant_property_wealth.png} 
%\end{figure}
%
\newpage

\section{Shadow rates and UMP identification}
\label{app:ssrs}

\begin{figure}
	\caption{The shadow rate as a proxy of Unconventional Monetary Policy}
	\label{fig:shadow_rate}
	\includegraphics[scale=0.5]{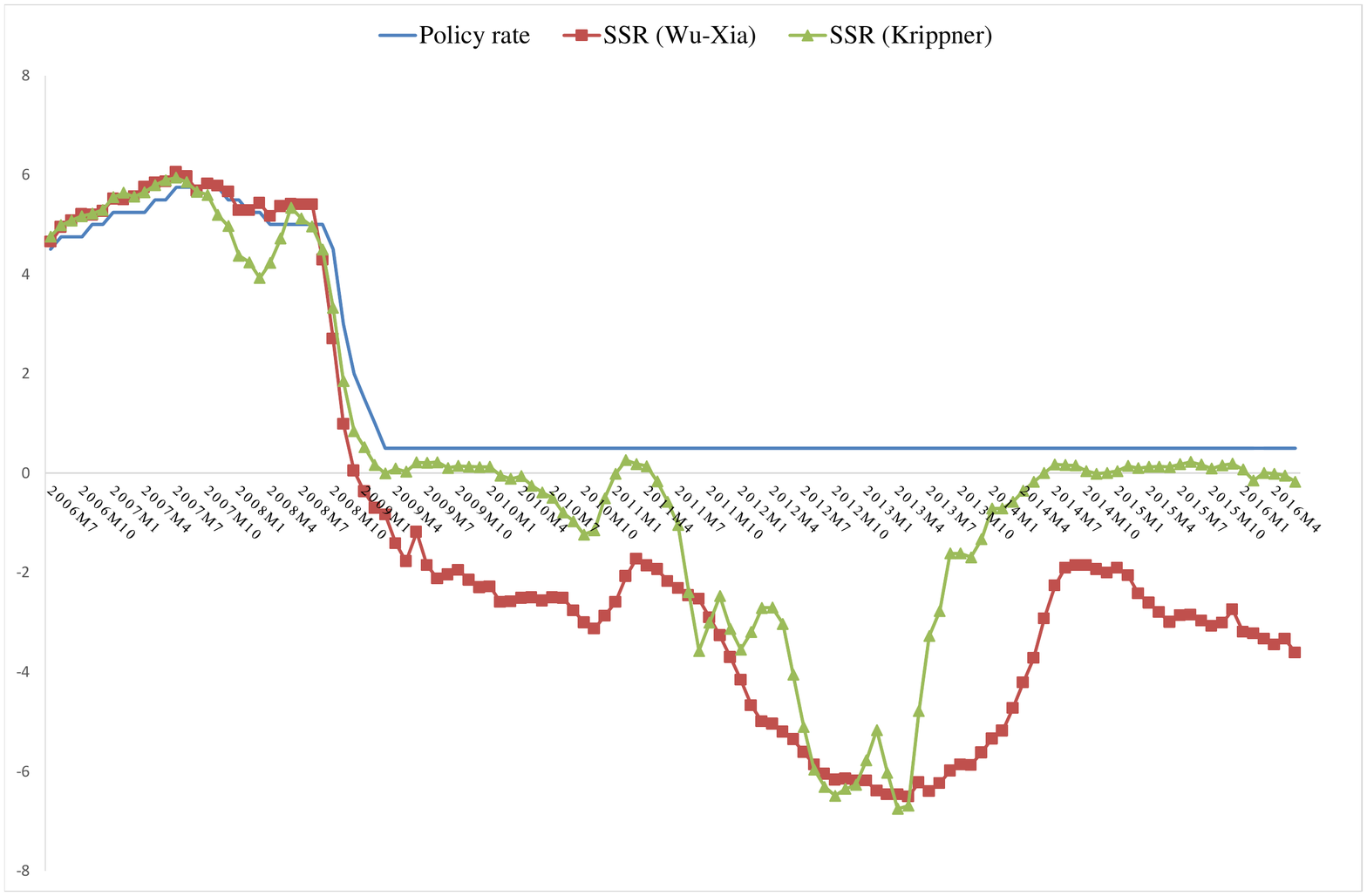} 
	\begin{figurenotes}[Source]
		\cite{wu2016measuring, krippner2014measuring},  FRED 
	\end{figurenotes}
\end{figure}

\subsection{Time varying VAR model}
\label{app:tvvar}

We use a TV-VAR model in order to conduct a counterfactual policy scenario to measure the impact of UMP on inequality as described in section \ref{sec:counterfactual}. This approach was proposed by \cite{baumeistera2013unconventional} who identified a shock to the 10 year government bond spread in order to examine the macroeconomic effects of a yield spread compression. 
Consider again the structural VAR in equation (1) but this time, following \cite{primiceri2005time}, the covariance matrix $\Sigma_t$ is decomposed as:

$${\Omega _t} = A_t^{ - 1}{H_t}{\left( {A_t^{ - 1}} \right)^{'}}$$

In our four VAR specification, the time-varying matrices ${H_t}$ and ${A_t}$ are defined as follows:
$$
{H_t} = \left[ {\begin{array}{*{20}{c}}
	{{h_{1,t}}}&0&0&0&0 \\ 
	0&{{h_{2,t}}}&0&0&0 \\ 
	0&0&{{h_{3,t}}}&0&0 \\ 
	0&0&0&{{h_{4,t}}}&0 
	\end{array}} \right] \text{	and	} {A_t} = \left[ {\begin{array}{*{20}{c}}
	1&0&0&0&0 \\ 
	{{a_{21,t}}}&1&0&0&0 \\ 
	{{a_{31,t}}}&{{a_{32,t}}}&1&0&0 \\ 
	{{a_{41,t}}}&{{a_{42,t}}}&{{a_{43,t}}}&1&0 
	\end{array}} \right]
$$

where ${H_t}$ is a diagonal matrix of the stochastic volatilities and ${A_t}$ is a lower triangular matrix which captures the contemporaneous interactions of the endogenous variables. Following \cite{primiceri2005time}, the elements of ${B_{i,t}},{h_{i,t}},{a_{ii,t}}$ are modeled as random walks. The advantage of this approach is that we allow for permanent shifts while we reduce the number of parameters to be estimated in a model which is already heavily parameterized. In particular, denoting ${h_t} = {\left[{{h_{1,t}},{h_{2,t}},{h_{3,t}},{h_{4,t}}} \right]'}$ and ${a_t} = {\left[ {{a_{21,t}},{a_{31,t}}, \ldots {a_{43,t}}} \right]'}$, we have that:

$ln{h_t} = ln{h_{t - 1}} + {n_t}$

${B_t} = {B_{t - 1}} + {\eta _t}$

${a_t} = {a_{t - 1}} + {\tau _t}$

where ${h_{i,t}}$ evolves as a geometric random walk and ${B_t}$, ${a_t}$ evolve as driftless random walks. We assume that the vector $\left[ {{\varepsilon _t},{\eta _t},{\tau _t},{\nu _t}} \right]{{\text{}}'}$  is distributed as:

$\left[ {\begin{array}{*{20}{c}}
	{{v_t}}\\
	{{\eta _t}}\\
	{{\tau _t}}\\
	{{n_t}}
	\end{array}} \right] \sim N\left( {0,V} \right), \text{ with } V = \left[ {\begin{array}{*{20}{c}}
	\Sigma &0&0&0 \\ 
	0&Q&0&0 \\ 
	0&0&S&0 \\ 
	0&0&0&Z 
	\end{array}} \right]$

\subsubsection{Estimation of the TV-VAR Model}
\label{app:tv_var}
The model is estimated using Bayesian methods. We describe in what follows the prior distributions as well as the estimation algorithm.
\subsubsection{Prior Distributions}

The initial conditions for the VAR coefficients $B_{0}$ are obtained via an OLS estimate of a fixed VAR using the first $T_{0}$ observations and then the prior distribution for B is defined as  ${B_0} \sim N\left[{{{\hat B}_{OLS,}},4x\hat V\left( {{{\hat B}_{OLS}}} \right)} \right]$. For the prior of h, let ${\hat \Sigma _{OLS}}$  be the estimated covariance matrix of  ${\nu _t}$  from the estimation of the time-invariant version of (1) and let K be the lower triangular Choleski factor under which  = $KK' = {\hat \Sigma _{OLS}}$. The prior is then defined as $ln{h_0}\sim N\left( {ln{\mu _0},10x{I_3}}\right){\text{}}$ where $\mu_0$ is a vector collecting the logarithms of the squared elements on the diagonal of $K$. For the prior of the off-diagonal elements of A, we set ${a_0}\sim N\left[ {{{\tilde a}_0},\tilde V\left( {{{\tilde a}_0}} \right)} \right]$  where ${\tilde a_0}$ are the off-diagonal elements of ${\hat \Sigma _{OLS}}$, where each row scaled by the corresponding element on the diagonal, while $\tilde V\left( {{{\tilde a}_0}} \right)$ is a diagonal matrix with each element (i,i) being 10 times the absolute value of the corresponding i-th element.

Regarding the prior distributions for the hyperparameters, the prior of $Q$ is assumed to be inverse Wishart distribution  
$Q\sim IW\left( {{T_0}\mathop Q\limits,{T_0}} \right)$. The scale parameter is equal to ${T_0}\mathop Q\limits$, where $\mathop Q\limits = \rho x{\hat \Sigma _{OLS}}$, and $\rho  = 0.0001$. The prior distribution of the elements of $S$ is assumed to be inverse Wishart ${S_i}\sim IW\left( {{S_{\mu {0_i}}},{S_{v{0_i}}}} \right)$  where $i$ indexes the blocks of $S$ where ${S_{\mu {0_i}}}$ is a diagonal matrix with the relevant elements of ${\tilde a_0}$ multiplied by ${10^{ - 3}}$  (see also \cite{mumtaz2017impact} who use this prior specification). Finally, for the variances of the stochastic volatility innovations, we set an inverse Gamma distribution for the elements of $Z$, $\sigma _i^2\sim IG\left( {{\sigma _{\mu 0}} = \frac{{0.0001}}{2},{\sigma _{\sigma 0}} = \frac{1}{2}} \right)$.

A Gibbs sampling algorithm is used to sample from the posterior distribution. The details of each conditional distribution are provided below.
\\
\\
\textit{1st step; drawing the coefficient states ${B_t}$}
\\
\\
Conditional on  ${A_t},{H_t}V$, the observation equation \ref{eq:1} is linear with Gaussian innovations and a known covariance matrix. Therefore, we draw ${B_t}$ using the \cite{carter1994} algorithm as follows. The conditional posterior distribution of  $p\left( {{B^T} \setminus {Y^T},{A^T},{H^T},V} \right)$ is written as $p\left( {{B^T} \setminus {Y^T},{A^T},{H^T},V} \right) = p\left( {{B_T} \setminus {Y^T},{A^T},{H^T},V} \right)\mathop \prod \limits_{t = 1}^{T - 1} p({B_t} \setminus {B_{t + 1}},{Y^T},{A^T},{H^T},V)$. The first term on the right hand side equation, i.e. the posterior distribution of ${B_t}$ is distributed as $p\left( {{B_T} \setminus {Y^T},{A^T},{H^T},V} \right) \sim N\left( {{B_{T|{\rm T}}},{P_{T|{\rm T}}}} \right)$. The second element, i.e. the posterior distribution of ${B_t}$, is distributed as $p\left( {{B_t} \setminus {B_{t + 1}},{Y^T},{A^T},{H^T},V} \right) \sim N\left( {{B_{t \setminus t + 1}},{P_{t \setminus t + 1}}} \right).{\text{}}$ The simulation proceeds as follows. First we use Kalman filter to draw ${B_{T|{\rm T}}},{P_{T|{\rm T}}}$ and then we proceed backwards in time by using ${B_{t \setminus t + 1}} = {B_{t \setminus t}} + {P_{t \setminus t}}P_{t \setminus t + 1}^{ - 1}\left( {{B_{t + 1}} - {B_t}} \right){\text{}}$ and ${P_{t \setminus t + 1}} = {P_{t \setminus t}} + {P_{t \setminus t}}P_{t \setminus t + 1}^{ - 1}{P_{t \setminus t}}$.
\\
\\
\textit{2nd step; draw the covariance states ${a_i}$} %% double check if this is i or t
\\
\\
Before describing this step, note that ${\nu_t}$, the VAR residuals, can be written as   with $var\left( {{\varepsilon _t}} \right) = {H_t}$. This is a system of linear equations with time varying coefficients and heteroskedasticity which has a known form. The jth equation of this system is given as  ${\nu _{jt}} =  - {a_{jt}}{\nu_{ - jt}} + {\varepsilon _{jt}}$ , where the subscript j denotes the jth column of ${\nu_t}$, while $-j$ denotes columns 1 to $j-1$. Note that this is a system of equations with time varying coefficients ${a_t}$. Following \cite{primiceri2005time}, we simplify the analysis by allowing the covariance of ${\tau _t}$, S, to be block diagonal, that is, the shocks to the jth equation ${\tau _{j,t}}$ are uncorrelated with those from other equations. Given this diagonal form, the elements of ${A_t}$ can be drawn by using the standard \cite{carter1994} algorithm.
\\
\\
\textit{3rd step; Draw the volatility states, ${h_t}$}
\\
\\
Following \cite{cogley2005drifts}, the diagonal elements of ${H_t}$ are sampled using a Metropolis Hastings algorithm. To see this, we write the following equation as before, ${A_t}{\nu _t} = {\varepsilon _t}$, where, $var\left( {{\varepsilon _t}} \right) = {H_t}$. Now, conditional on ${B_t}$ and ${A_t}$, the distribution of ${h_{it}}$ is given by:

$f({h_{it}}/{h_{it - 1}}$${h_{i,t + 1,}}{\varepsilon _{it}}) = f\left( {{\varepsilon _{it}} \setminus {h_{it}}} \right)xf({h_{it}} \setminus {h_{it - 1}})xf\left( {{h_{it + 1}} \setminus {h_{it}}} \right)$

$ = h_{it}^{ - 0.5}exp\left( {\frac{{ - \varepsilon _{it}^2}}{{2{h_{it}}}}} \right)xh_{it}^{ - 1}\exp \left( {\frac{{ - {{\left( {ln{h_{it}} - \mu } \right)}^2}}}{{2{\sigma _{{h_i}}}}}} \right)$

where $\mu$ and ${\sigma_{{h_i}}}$ denote the mean and variance of the log-normal density $h_{it}^{ - 1}\exp \left( {\frac{{ - {{\left( {ln{h_{it}} - \mu } \right)}^2}}}{{2{\sigma _{{h_i}}}}}} \right)$. Following \cite{jacquier2002bayesian}, we use this log normal density as the candidate generating density with the acceptance probability defined as the ratio of the conditional likelihood $h_{it}^{ - 0.5}exp\left( {\frac{{ - \varepsilon _{it}^2}}{{2{h_{it}}}}} \right)$ at the old and the new draw. This algorithm is applied at each period in the sample to deliver a draw of the stochastic volatilities.
\\
\\
\textit{4th  step, draw the hyperparameters, $Q, S, Z$}
\\
\\
Conditional on $B_t, A_t, H_t $, we sample the hyperparameters as follows: Q is sampled from the inverse Wishart distribution using the scale matrix $\eta _t'{\eta _t} + {Q_0}$ and degrees of freedom $T + T_0$. Next, S is sampled from 
the inverse Gamma distribution with scale parameter $\tau _t'{\tau _t} + {S_i}$ and degrees of freedom $T+T_0$. Last, we 
draw the elements of Z from its inverse Wishart distribution with scale parameter   $\frac{{{{\left( {ln{h_{it}} - ln{h_{it - 1}}} \right)}'}\left( {ln{h_{it}} - ln{h_{it - 1}}} \right) + {\sigma _{\mu 0}}}}{2}$ and degrees of freedom,  $\frac{{T + {\sigma _{\sigma 0}}}}{2}$. The algorithm is run for 100,000 iterations discarding the initial 60,000 as burn-in sample.

\subsection{Threshold VAR}
\label{app:threshold_var}

The TVAR is defined as:

\begin{equation}
\label{eq:tvar}
\begin{split}
{y_t} = {c_1} + \mathop \sum \limits_{j = 1}^p {B_{1j}}{y_{t - j}} + {v_t} 
,\quad {v_t} \sim N\left( {0,\Sigma_1} \right),  \text{if	} {Y_i}_{t-d} <= {Y^*} \\
{y_t} = {c_2} + \mathop \sum \limits_{j = 1}^p {B_{2j}}{y_{t - j}} + {\epsilon_t}
,\quad {\epsilon_t} \sim N\left( {0,\Sigma_2} \right),  \text{if	} {Y_i}_{t-d} > {Y^*}
\end{split}
\end{equation}

where ${Y_i}_{t-d}$ is the threshold variable which in our case is the shadow rate, $d$ is the time lag that is assumed to be known and $Y*$  is the threshold level. Based on standard information criteria, the specification that we follow is a one lag VAR with the threshold variable delayed by two periods. 

% the following par. is not needed but please do not delete it
%We estimate the TVAR depicted in \ref{eq:tvar} by performing a Gibbs algorithm to obtain the parameters $c_i, B_i, Y^*, \Sigma_i$,  where $i=1,2$, depending on which regime the economy lies. The Gibbs algorithm is described in the appendix but the main intuition is straightforward. First, we sample the VAR parameters $b_i=(c_i,B_i)$ from a conditional posterior normal distribution. Given a draw of $b_i$, $\Sigma_i$ is drawn from a conditional inverse-Wishart distribution. Finally, given the values for coefficients and covariances, we sample the threshold parameter $Y^*$ by a using a Metropolis Hastings (MH) algorithm within the Gibbs algorithm.  

\subsubsection{Estimation of the TVAR}
Following \cite{bandura2010} we introduce a dummy observation prior for the VAR parameters  ${{\text{b}}_{\text{i}}} = \left\{ {{{\text{c}}_{\text{i}}},{\text{}}{{\text{B}}_{\text{i}}}} \right\}$, where i=1,2. The prior means are chosen as the OLS estimates of the coefficients of an AR(1) regression estimated for each endogenous variable using a training sample. As is standard in the literature, we set the overall prior tightness $\lambda= 0.1$. Next, we assume that the prior of   follows the normal distribution with $p\left( {{Y^*}} \right)\sim N\left( {{{\bar Y}^*},{\sigma _{{Y^{*}}}}} \right).$ We follow \cite{blake2012applied} by using the mean of the threshold variable as ${\bar Y^*}$ and the variance of the series as ${\sigma _{{Y^*}}}$. 

 For simplicity, denoting the right hand side variables of the TVAR as $X_i$ , we can write the conditional posterior distribution of $b_i$ that is standard and follows the normal distribution as:

\begin{equation}
\label{eq:dist_tv_1}
H\left( {{b_i}\backslash {\Sigma _{i,}}{y_i},{Y^*}} \right)\sim N\left[ {vec\left( {B_i^*} \right),{\Sigma _i} \otimes {{\left( {X_i^{*'}X_i^*} \right)}^{ - 1}}} \right]
\end{equation}
where $B_i^* = {\left( {X_i^{*'}X_i^*} \right)^{ - 1}}X_i^{*'}y_i^*)$ and $y_i^*,{\text{X}}_{\text{i}}^{\text{*}}$    denote the transformed data in regime i augmented with the dummy observations that define the prior for the left and the right hand side of the TVAR respectively. The conditional posterior distribution of $\Sigma_i$  is given by the inverse Wishart distribution:         

\begin{equation}
\label{eq:dist_tv_2}
H\left( {{\Sigma _i}\backslash {b_{i,}}{y_i},{Y^*}} \right)\sim IW\left( {S_i^*,T_i^*} \right){\text{}}
\end{equation}
where $S_i^* = {\left( {y_i^* - X_i^**{b_i}} \right)'}\left( {y_i^* - X_i^**{b_i}} \right){\text{}}$ and $T_i^*$  denotes the number of rows in ${Y^*}$.

	The Gibbs sampler cycles through the following steps: (i) the parameters   are sampled in each regime according to the conditional posterior distributions \ref{eq:dist_tv_1} and \ref{eq:dist_tv_2}. Then, given the values for coefficients and covariances, (ii) we sample the threshold value, ${Y^*}$ by using a Metropolis Hastings random walk algorithm as follows.

 We draw a new value of the threshold from the random walk process: ${\rm{Y}}_{{\rm{new}}}^{\rm{*}}= {Y}_{{\text{old}}}^{\text{*}} + {{\Psi }^{1/2}}{e}$, ${\text{e}}\sim {\text{N}}\left( {0,{\Sigma }} \right)$ , where ${{\rm{\Psi }}^{1/2}}$ is a scaling factor that is chosen so as to ensure that the acceptance rate is in the 20–40\% interval. Next, we compute the acceptance probability:
\begin{equation}
a = \frac{{F\left( {y\backslash {b_i},{\Sigma _i},Y_{new}^*} \right)p\left( {Y_{new}^*} \right)}}{{F\left( {y\backslash {b_i},{\Sigma _i},Y_{old}^*} \right)p\left( {Y_{old}^*} \right)}}
\end{equation}
where $F\left( {y\backslash {b_i},{\Sigma _i},Y_{new}^*} \right)$ is the likelihood of the parameters as the product of the likelihoods in the two regimes. The log likelihood in each regime is defined as: $lnF = \left( {\frac{T}{2}} \right)log\left| {\Sigma _i^{ - 1}} \right| - 0.5\mathop \sum \limits_{t = 1}^T \left[ {({Y_{i,t}} - {X_{i,t}}{b_i}} \right.{)'}\Sigma _i^{ - 1}\left. {({Y_{i,t}} - {X_{i,t}}{b_i})} \right]{\text{}}.{\text{}}$We then draw $u\sim U\left( {0,1} \right)$. If $u < a$, accept $Y_{new}^*$ else retain $Y_{old}^*$. We run 100,000 draws and discard the first 60,000 to ensure convergence.

% \newpage

% \section{Robustness Checks}
% \label{app:robustness}

% \begin{figure}[!htbp]
% 	\caption{The impulse response of the Gini coefficient to a monetary policy shock using BoE total assets innovations an UMP shock }
% 	\label{fig:robust_assets}
% 	\includegraphics[scale=.7]{robustness_total_assets.png}
% 	\begin{figurenotes}[Note]
% The vertical axis of each plot shows the response in percent. Time intervals on the x-axis are months. The red line is the median estimate and the yellow shaded area is the error band. 
% 	\end{figurenotes}
% \end{figure}

% \begin{figure}[!htbp]
% 	\caption{The impulse response of the Palma ratio to a monetary policy shock}
% 	\label{fig:robust_palma}
% 	\includegraphics[scale=.7]{robustness_palma.png}
% 	\begin{figurenotes}[Note]
% 		The vertical axis of each plot shows the response in percent. Time intervals on the x-axis are months. The red line is the median estimate and the yellow shaded area is the error band. 
% 	\end{figurenotes}
% \end{figure}

\newpage
\section{Additional estimates}
\label{sec:additional}

\begin{figure}
	\caption{Impulse responses of net financial wealth quantiles}
		\label{fig:financial_shares}
	\includegraphics[scale=.6]{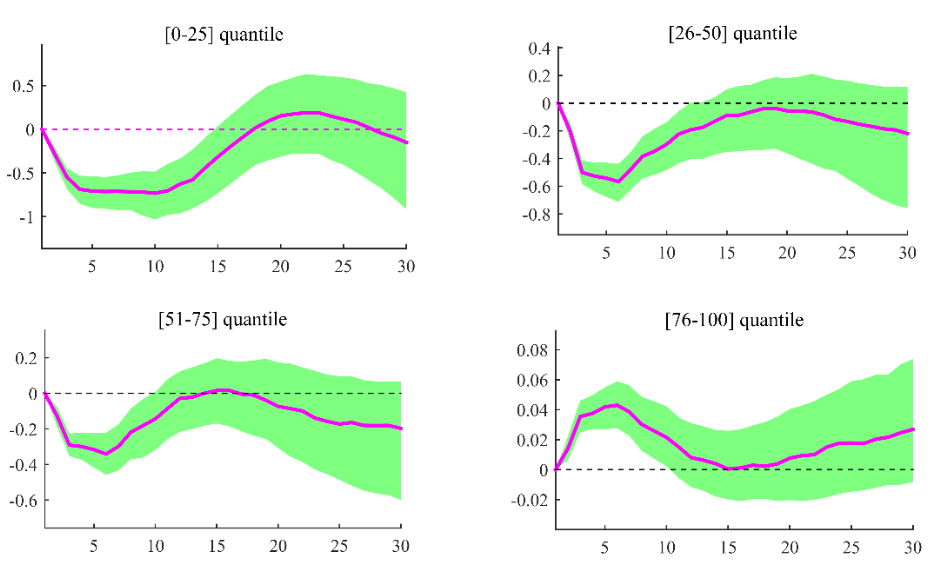} 
			\begin{figurenotes}[Note]
		The vertical axis of each plot shows the response in percent. Time intervals on the x-axis are months. The pink line is the median estimate and the green shaded area is the error band. 
	\end{figurenotes}
\end{figure}

\begin{figure}
	\caption{Impulse responses of net housing wealth quantiles}
		\label{fig:housing_shares}
	\includegraphics[scale=.4]{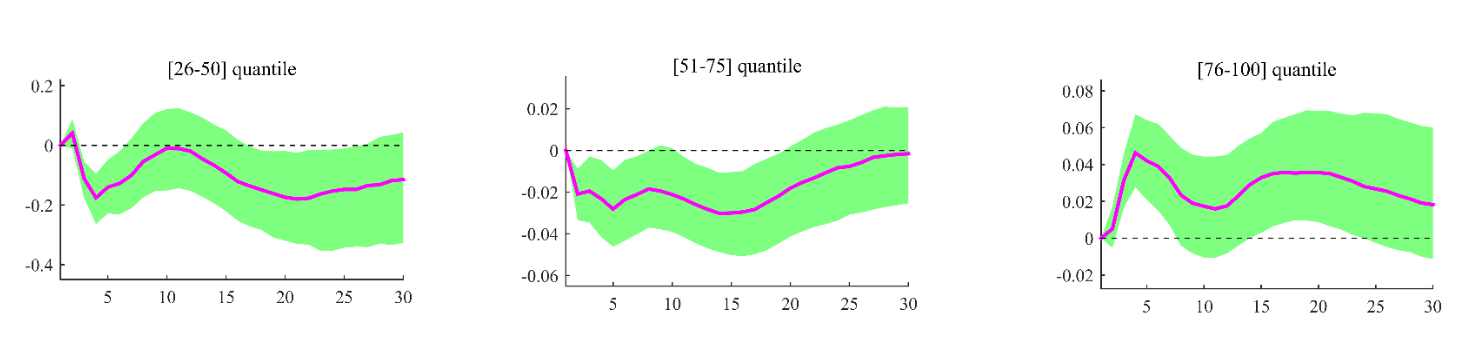} 
			\begin{figurenotes}[Note]
		The vertical axis of each plot shows the response in percent. Time intervals on the x-axis are months. The pink line is the median estimate and the green shaded area is the error band. 
	\end{figurenotes}
\end{figure}

\end{document}